# A Two-dimensional HLLC Riemann Solver for Conservation Laws: Application to Euler and Magnetohydrodynamic Flows

By

Dinshaw S. Balsara (dbalsara@nd.edu) Physics Department, Univ. of Notre Dame

**Abstract**

In this paper we present a genuinely two-dimensional HLLC Riemann solver. On logically rectangular meshes, it accepts four input states that come together at an edge and outputs the multi-dimensionally upwinded fluxes in both directions. This work builds on, and improves, our prior work on two-dimensional HLL Riemann solvers. The HLL Riemann solver presented here achieves its stabilization by introducing a constant state in the region of strong interaction, where four one-dimensional Riemann problems interact vigorously with one another. A robust version of the HLL Riemann solver is presented here along with a strategy for introducing sub-structure in the strongly-interacting state. Introducing sub-structure turns the two-dimensional HLL Riemann solver into a two-dimensional HLLC Riemann solver. The sub-structure that we introduce represents a contact discontinuity which can be oriented in any direction relative to the mesh.

The Riemann solver presented here is general and can work with any system of conservation laws. We also present a second order accurate Godunov scheme that works in three dimensions and is entirely based on the present multidimensional HLLC Riemann solver technology. The methods presented are cost-competitive with traditional higher order Godunov schemes.

The two-dimensional HLLC Riemann solver is shown to work robustly for Euler and Magnetohydrodynamic (MHD) flows. Several stringent test problems are presented to show that the inclusion of genuinely multidimensional effects into higher order Godunov schemes indeed produces some very compelling advantages. For two dimensional problems, we were routinely able to run simulations with CFL numbers of ~0.7, with some two-dimensional simulations capable of reaching higher CFL numbers. For three dimensional problems, CFL numbers as high as ~0.6 were found to be stable. We show that on resolution-starved meshes, the scheme presented here outperforms unsplit second order Godunov schemes that are based on conventional one-dimensional Riemann solver technology. Strong discontinuities are shown to propagate very isotropically using the methods presented here. The present Riemann solver provides an elegant resolution to the problem of obtaining multi-dimensionally upwinded electric fields in MHD without resorting to a doubling of the dissipation in each dimension.

## I) Introduction

Ever since the introduction of Godunov schemes and their higher order extensions (Godunov [30], van Leer [49]), Riemann solvers have been viewed as an essential component in the design of robust schemes for computational fluid dynamics (CFD). Early advances in CFD, therefore, gave rise to several very sophisticated Riemann solvers. The exact Riemann solver of van Leer [49] and the two-shock Riemann solver (Colella [17], Colella & Woodward [19]) were two early attempts to represent shock jumps exactly. Riemann solvers were also designed by Osher and Solomon [38] by relying on rarefaction fans instead of shock waves. The linearized Riemann solver by Roe [40] and the HLLE Riemann solver (Harten, Lax & van Leer [32], Einfeldt [23]) were early attempts to retain an essential wave model while discarding flow features within the Riemann problem that do not contribute to the numerical flux. The HLLC Riemann solvers (Einfeldt *et al.* [24], Toro, Spruce and Speares [48], Batten *et al.* [11]) took this idea one step further by retaining a wave model that is true to the physics of the problem while avoiding the need to linearize the hyperbolic system. All the Riemann solvers mentioned in this paragraph were strictly one-dimensional so that multidimensional flow features were decomposed into discontinuities that propagate along the dominant directions of the mesh. Mesh imprinting, where the same flow features propagate at different speeds in different directions on a computational mesh, is an inevitable consequence of this approach. A CFL number that decreases as the dimensionality of the problem is increased is another unavoidable result.

Roe [41] and Rumsey, van Leer & Roe [43] made an early attempt to introduce multidimensionality into the Riemann solvers, but the wave models they postulated either proved analytically intractable or unworkable. Consequently, some practitioners have attempted to use the one-dimensional Riemann solvers in very intricate combinations in order to achieve multidimensional upwinding (Colella [18], Saltzman [46], LeVeque [34]). This form of multidimensional upwinding did indeed enable the design of schemes that operate with an increased Courant number, even though it sometimes came at the expense of solving a rather large number of one-dimensional Riemann problems.

Early successes in formulating a genuinely multi-dimensional Riemann solver for hydrodynamics emerged in the work of Abgrall [1], [2] and were followed up by the work of other practitioners (Fey [26], [27], Gilquin, Laurens & Rosier [29], Brio, Zakharian & Webb [14]). These Riemann solvers were multidimensional extensions of the linearized Riemann solver of Roe [40]. They were specific to the Euler equations, required elaborate constructions and were hard to extend to more general systems of conservation laws. Wendroff [50] formulated a two-dimensional HLL Riemann solver, but did not present an easy to implement formulation, nor did he provide a positivity demonstration. Working on logically rectangular meshes, Balsara [7] presented a two-dimensional HLL Riemann solver with simple closed form expressions for the fluxes that were easy to implement. By showing that the two-dimensional Riemann solver was capable of treating Euler and MHD flows, he illustrated that it was flexible enough to handle several different kinds of conservation laws. He also demonstrated the positivity of the Riemann solver for the Euler equations. In this work we present an even more robust version of the same HLL Riemann solver. It is more robust because it incorporates effects from the transverse fluxes in a more intricate fashion.

When formulating two-dimensional Riemann solvers, it helps to think of the two space directions and the temporal direction as a single three-dimensional entity. The three essential ideas in the design of such multi-dimensional HLL Riemann solvers are:

1) Using one-dimensional Riemann solvers along each of the four one-dimensional jumps in the flow, we obtain the maximal wave speeds in each of the principal directions of the logically rectangular mesh. These maximal wave speeds can be used to build a wave model for the two-dimensional Riemann solver. (Unstructured meshes will be treated in a separate paper.)

2) Realize that there is a central region where the four resolved states from the one-dimensional Riemann problems interact strongly with one another. By observing the solutions of several two-dimensional Riemann problems from Schulz-Rinne, Collins and Glaz [47], we can realize that there is indeed such a region of strong interaction and that its extent is bounded by the maximal speeds of each of the one-dimensional Riemann solvers. We refer to this state as the strongly-interacting state of the multi-dimensional Riemann solver. It is indeed the logical analogue of the subsonic (or resolved) state in the one-dimensional HLL Riemann solver. (For those who would like to preview what the two-dimensional Riemann problem looks like, we display some examples in Sub-section V.a.2.) As with the one-dimensional HLL Riemann solver, this state is taken to be a constant. By taking it to be a constant, we provide sufficient dissipation in situations where this strongly-interacting state straddles the time-axis. For a one-dimensional HLL Riemann solver, this resolved state is obtained via a two-dimensional integration in space-time. In an analogous fashion, the strongly-interacting state in the two-dimensional Riemann solver can be obtained via a three-dimensional space-time integration.

3) Recall that the subsonic flux for the one-dimensional HLL Riemann solver is obtained by integrating over a smaller sub-domain in a two-dimensional space-time. This sub-domain is chosen so as to include the time axis as one of its boundaries because integration along that boundary yields the desired numerical flux. In an entirely analogous fashion, when the strongly-interacting state in the two-dimensional Riemann solver straddles the time axis, the numerical fluxes have to be obtained by intricate space-time integrations. In such a situation, the two numerical fluxes in the strongly-interacting region of the two-dimensional HLL Riemann solver are obtained by integrating over two smaller sub-domains in a three-dimensional space-time. As in the one-dimensional case, these two sub-domains are chosen so that their boundaries pass through the time axis. Each of these two integrations yields one flux component. In this work we present a two-dimensional HLL Riemann solver that is significantly more robust than the one in Balsara [7] while retaining the advantage of having simple closed form expressions for the strongly-interacting state and fluxes.

The one-dimensional HLLC Riemann solver is obtained by introducing sub-structure associated with a contact discontinuity into the subsonic state of the one-dimensional HLL Riemann solver. The analogy extends to the two-dimensional HLLC Riemann solver. A study of the two-dimensional Riemann problems from Schulz-Rinne, Collins and Glaz [47] shows that the region of strong interaction can give rise to a rich array of flow structures. However, a majority of those flow structured consisted of contact discontinuities; consequently, these are the flow sub-structures that we must try to capture. Furthermore, we point out that Roe [41] attempted to endow his multi-dimensional Riemann solver with so much sub-structure that the mathematics became intractable. We, therefore, take a minimalist approach and only allow the strongly-interacting state from the two-dimensional HLL Riemann solver to be partitioned into two sub-domains which represent the two sides of a contact discontinuity. The speed with which the contact discontinuity propagates in two dimensions is determined by the two-dimensional HLL Riemann solver. This is similar to the way the one-dimensional HLL Riemann solver provides the speed for the contact discontinuity in the one-dimensional HLLC Riemann solver. The size of the jump in flow variables across that contact discontinuity will be determined by conservation principles, analogously to the way it is done in the one-dimensional HLLC Riemann solver. However, following Roe [41], we realize that the orientation of the contact discontinuity should be provided by examining the larger density variation in the flow. To appreciate this from another viewpoint, realize that the multi-dimensional Riemann solver operates at only one point; hence it cannot discern the direction of the contact discontinuity from the four states that form its inputs. In this paper we present a two-dimensional HLLC Riemann solver that is constructed in keeping with the principles described in this paragraph.

The HLL and HLLC Riemann solvers presented here are very general and can be applied to several systems of conservation laws. To demonstrate this versatility, we design a second order accurate scheme for Euler and MHD flows that is based entirely on the two-dimensional HLLC Riemann solver technology. For two-dimensional problems, the scheme is theoretically stable up to a CFL number of unity; though an even-odd decoupling (discussed later) requires us to use somewhat lower CFL numbers. Numerical experimentation shows that the same scheme applied to three-dimensional problems achieves a stable CFL number as high as ~0.6, though we have no rigorous proof to that effect. The two-dimensional HLLC Riemann solver adds little to the computational complexity of the two-dimensional HLL Riemann solver, but it goes a long way in lowering dissipation. Rigorous tests show that our second order scheme enables strong discontinuities to propagate much more isotropically than any second order Godunov scheme that is built on one-dimensional Riemann solver technology. Indeed, the second order scheme presented here has wave propagation characteristics that are more isotropic than many third, fourth and fifth order Godunov schemes that we have had experience with. In a subsequent paper we will show how this multi-dimensional Riemann solver technology can be melded into ADER-WENO schemes that go beyond second order. (ADER stands for Arbitrary DERivatives in space and time; WENO stands for weighted essentially non-oscillatory.)

MHD provides an interesting example of a hyperbolic system that is larger and admits more complicated wave structures than the Euler system. One-dimensional linearized Riemann solvers for numerical MHD have been designed (Roe & Balsara [42], Cargo and Gallice [15], Balsara [3]). HLLC Riemann solvers, capable of capturing mesh-aligned contact discontinuities, have been presented by Gurski [31] and Li [35]. Miyoshi and Kusano [37] drew on Gurski's work to design an HLLD Riemann solver for MHD. In addition to contact discontinuities, the HLLD Riemann solver was also capable of capturing mesh-aligned Alfven waves. In this paper we wish to catalogue the performance of the two-dimensional HLLC Riemann solver for MHD problems.

As shown in Balsara [7], multi-dimensional Riemann solvers play an essential role in numerical MHD. The magnetic field in the MHD system satisfies the property that it remains divergence-free for all time. Brackbill & Barnes [12] have shown that violating the divergence-free aspect of the magnetic field leads to unphysical plasma transport orthogonal to the magnetic field. One possible resolution is to formulate constrained transport schemes (Brecht *et al.* [13], DeVore [22], Evans & Hawley [25]) which collocate magnetic fields at zone centers and use edge-centered electric fields for their divergence-free update. This collocation is mandated by Faraday's law which states that the time rate of update of the magnetic field is proportional to the curl of the electric field. Notice that the curl operator is best discretized using Stokes law instead of Gauss law. Other approaches, all of which involve modifying the MHD equations, include the fix by Powell [39] and the GLM method of Dedner *et al.* [21]. Soon after the advent of higher order Godunov schemes for MHD, Dai & Woodward [20], Ryu *et al.* [45] and Balsara & Spicer [10] formulated higher order Godunov methods that kept the magnetic field divergence-free. The essential idea in Balsara & Spicer [10], which is now widely used in other schemes, consisted of realizing that there is a dualism between the upwinded fluxes (when the MHD system is written out as a conservation law) and the electric fields (that are needed for updating the magnetic fields according to Faraday's law). Balsara & Spicer [10] also pointed out that the electric fields have to be obtained via a multi-dimensional upwinding and tried to build multidimensional effects within the context of the one-dimensional Riemann solver technology that was available at that time. Recent work has tried to increase the amount of dissipation from one-dimensional Riemann solvers to stabilize magnetic fields that propagate in directions that are not grid aligned (Londrillo and DelZanna [36], Gardiner & Stone [28]). If this dissipation doubling is viewed as an expedient, it might be comprehensible; but viewed as a properly formulated numerical method, we fail to comprehend it. In this paper, we extend the research that was begun in Balsara [7] and show that such a doubling of the dissipation is completely unnecessary when the multi-dimensional HLLC Riemann solver developed here is used to obtain the electric field.

The plan of this paper is as follows. Section II presents the multidimensional HLL and HLLC Riemann solvers. This is done in the context of Euler flow. Section III extends the work to MHD. Section IV presents the numerical scheme. Section V presents numerical results. Section VI provides our conclusions.

## II) Derivation of the Two-Dimensional HLLC Riemann Solver

This section is split into two sub-sections. Sub-section II.a presents the derivation of a robust HLL Riemann solver in two dimensions. Sub-section II.b extends this work to include the HLLC Riemann solver. The presentation in this section is done within the context of the Euler system.

## II.a) Derivation of an Improved HLL Riemann Solver in Two Dimensions

The two-dimensional HLLC Riemann Solver is built by drawing on the two-dimensional HLL Riemann Solver of Balsara [7] in quite the same way that the one-dimensional HLLC Riemann Solver of Toro, Spruce and Speares [48] is built by drawing on the one-dimensional HLL Riemann Solver of Harten, Lax and van Leer [32]. However, we first make one small but important modification to the HLL Riemann solver of Balsara [7] which, in-fact, further adds to its stability.

Consider the *N*-component vector of conserved variables $\mathbf{U}$ that evolves according to the conservation law

$$\partial_t \mathbf{U} + \partial_x \mathbf{F} + \partial_y \mathbf{G} = 0 \qquad (1)$$

where $\mathbf{F}$ and $\mathbf{G}$ are the fluxes in the *x*- and *y*-directions. For the Euler equations, the five components of $\mathbf{U}$ are listed in order as the density, *x*-, *y*- and *z*-momentum and energy density. The two-dimensional Riemann solver is formulated on a Cartesian, or logically rectangular mesh. Fig. 1a shows how four zones come together at an edge "*O*" where the multidimensional

Riemann solver is formulated. The four states that come together at that edge are given by $\mathbf{U}_{RU}$, $\mathbf{U}_{LU}$, $\mathbf{U}_{LD}$, and $\mathbf{U}_{RD}$. Such situations, where four fluid states come together at an edge, have been studied by Schulz-Rinne, Collins and Glaz [47]. Their study of the multidimensional Riemann problem was carried out using conventional higher order Godunov methods on very well-resolved meshes. They give rise to four one-dimensional Riemann problems along with a region of strong interaction where all the one-dimensional Riemann problems interact strongly with one another. Thus we have one x-directional Riemann problem between $\mathbf{U}_{RU}$ and $\mathbf{U}_{LU}$ and another x-directional Riemann problem between $\mathbf{U}_{RD}$ and $\mathbf{U}_{LD}$. We also have one y-directional Riemann problem between $\mathbf{U}_{RU}$ and $\mathbf{U}_{RD}$ and another y-directional Riemann problem between $\mathbf{U}_{LU}$ and $\mathbf{U}_{LD}$. In the region of strong interaction, one can have very complicated self-similar fluid flow structures. It would be futile to resolve all those structures in a two-dimensional Riemann problem. When formulating a multidimensional HLL Riemann solver, we therefore average over the sub-structure in the region of strong interaction by assuming that it is all lumped into a single constant state. The extent of that state in space and time is then determined by the wave model that we adopt. We describe the wave model in the next paragraph.

In a one-dimensional Riemann solver, we wish to pick out the extremal speeds in the direction of wave propagation. In a two-dimensional Riemann solver, we wish to identify the maximal speeds in both the principal directions. Let $\lambda_x^1(\mathbf{U}_{RU})$ and $\lambda_x^N(\mathbf{U}_{RU})$ denote the smallest and largest x-directional wave speeds respectively in the state $\mathbf{U}_{RU}$, with corresponding definitions for the other states. Let $\bar{\lambda}_x^1(\mathbf{U}_{LU}, \mathbf{U}_{RU})$ and $\bar{\lambda}_x^N(\mathbf{U}_{LU}, \mathbf{U}_{RU})$ be the smallest and largest x-directional wave speeds from a linearized Riemann solver that is applied between the states $\mathbf{U}_{LU}$ and $\mathbf{U}_{RU}$, with similar definitions for the other pairs of states. Make similar definitions for the y-direction. Within the upper face of Fig. 1a, the fastest right- and left-going wave speeds are denoted by $S_R^U$ and $S_L^U$ respectively, and they are given by

$$S_R^U = \max\left(\lambda_x^N(\mathbf{U}_{RU}), \bar{\lambda}_x^N(\mathbf{U}_{LU}, \mathbf{U}_{RU})\right) \quad ; \quad S_L^U = \min\left(\lambda_x^1(\mathbf{U}_{LU}), \bar{\lambda}_x^1(\mathbf{U}_{LU}, \mathbf{U}_{RU})\right) \qquad (2)$$

Within the lower face of Fig. 1a, the fastest right- and left-going signal speeds are denoted by $S_R^D$ and $S_L^D$ respectively, and they are given by

$$S_R^D = \max\left(\lambda_x^N(\mathbf{U}_{RD}), \bar{\lambda}_x^N(\mathbf{U}_{LD}, \mathbf{U}_{RD})\right) \quad ; \quad S_L^D = \min\left(\lambda_x^1(\mathbf{U}_{LD}), \bar{\lambda}_x^1(\mathbf{U}_{LD}, \mathbf{U}_{RD})\right) \qquad (3)$$

Taken together, the above two sets of equations enable us to obtain the maximal right- and left-going wave speeds that bound the strongly-interacting state, denoted by $S_R$ and $S_L$ respectively, as

$$S_R = \max\left(S_R^U, S_R^D\right) \quad ; \quad S_L = \min\left(S_L^U, S_L^D\right) \qquad (4)$$

In an analogous fashion, the fastest upward- and downward-going wave speeds in the right face of Fig. 1a are denoted by $S_U^R$ and $S_D^R$ respectively, and they are given by

$$S_U^R = \max\left(\lambda_y^N(\mathbf{U}_{RU}), \bar{\lambda}_y^N(\mathbf{U}_{RD}, \mathbf{U}_{RU})\right) \quad ; \quad S_D^R = \min\left(\lambda_y^1(\mathbf{U}_{RD}), \bar{\lambda}_y^1(\mathbf{U}_{RD}, \mathbf{U}_{RU})\right) \tag{5}$$

In the left face of Fig. 1a, the fastest upward- and downward-going wave speeds are given by

$$S_U^L = \max\left(\lambda_y^N(\mathbf{U}_{LU}), \bar{\lambda}_y^N(\mathbf{U}_{LD}, \mathbf{U}_{LU})\right) \quad ; \quad S_D^L = \min\left(\lambda_y^1(\mathbf{U}_{LD}), \bar{\lambda}_y^1(\mathbf{U}_{LD}, \mathbf{U}_{LU})\right) \tag{6}$$

As before, the above two sets of equations enable us to obtain the maximal upward- and downward-going wave speeds that bound the strongly-interacting state, denoted by $S_U$ and $S_D$ respectively, as

$$S_U = \max\left(S_U^R, S_U^L\right) \quad ; \quad S_D = \min\left(S_D^R, S_D^L\right) \tag{7}$$

Using the wave speeds that were defined above, the wave model for the multi-dimensional Riemann problem can be specified.

Fig. 2a shows the three-dimensional space-time diagram associated with the wave model. I.e., time forms the third dimension. Specifically, we show the outer surface of a rectangular prism whose base is made up of the rectangle *RQMN* in Fig. 1a. The rectangular prism contains the region of strong interaction where all of the four one-dimensional Riemann problems interact with each other by the time *T*. We will provide further details on the region of strong interaction in the next paragraph. The figure also shows the shaded regions in space-time that are overrun by the one-dimensional HLL Riemann problems in the right and lower faces along with their specific wave speeds. We will now use Fig. 2a to define many of the fluxes and states that are useful in the construction of the multi-dimensional HLL Riemann solver. Thus, between the states $\mathbf{U}_{LD}$ and $\mathbf{U}_{RD}$ for instance, we have an *x*-directional HLL Riemann solver. This yields right- and left-going waves in the lower face, given by $S_R^D$ and $S_L^D$ respectively. These two waves form the boundaries of a resolved state given by $\mathbf{U}_D^*$. The above-mentioned *x*-directional HLL Riemann solver also produces a numerical *x*-flux, which we denote as $\mathbf{F}_D^{HLL}$. We can use the resolved state $\mathbf{U}_D^*$ and its associated resolved flux $\mathbf{F}_D^*$ to build the *y*-flux $\mathbf{G}_D^*$ in a way that is sketched out in the next paragraph. Likewise, in the right face we can see from Fig. 2a that between the states $\mathbf{U}_{RD}$ and $\mathbf{U}_{RU}$ we have a *y*-directional HLL Riemann solver. This yields upper- and lower-going waves given by $S_U^R$ and $S_D^R$ respectively. These two waves form the boundaries of the resolved state given by $\mathbf{U}_R^*$. The above-mentioned *y*-directional HLL Riemann solver also produces a numerical *y*-flux, which we denote as $\mathbf{G}_R^{HLL}$. Similarly, we can use the resolved state $\mathbf{U}_R^*$ and its associated resolved flux $\mathbf{G}_R^*$ to build the *x*-flux $\mathbf{F}_R^*$. Fluxes in the two faces that are not shown in Fig. 2a can be named and built similarly.

The construction of $\mathbf{G}_D^*$ from $\mathbf{U}_D^*$ and $\mathbf{F}_D^*$ is detailed as follows for the Euler equations. The pressure that is needed in the evaluation of $\mathbf{G}_D^*$ is obtained from the second components of $\mathbf{U}_D^*$ and $\mathbf{F}_D^*$. The fifth component of $\mathbf{U}_D^*$ provides the total energy density. The *x*-velocity is given by the ratio of the second to the first component of $\mathbf{U}_D^*$, with other velocity components obtained similarly. This information enables us to construct $\mathbf{G}_D^*$.

The multidimensionality of the HLL Riemann solver stems from the fact that the strongly-interacting state of the two-dimensional Riemann problem is indeed given by $\mathbf{U}^*$ in Fig. 2a. To understand the wave model better, imagine removing the two *x-t* and *y-t* boundaries of Fig. 2a. This would reveal just the strongly-interacting state $\mathbf{U}^*$ of the two-dimensional Riemann problem. This is shown in Fig. 3a, which shows the inverted pyramidal volume in space-time covered by the strongly-interacting state $\mathbf{U}^*$. Please notice that the wave model and states in our multi-dimensional Riemann solver are entirely self-similar, which is exactly what we expect from a Riemann solver.

Obtaining the strongly-interacting state $\mathbf{U}^*$ requires an integration of the conservation law over the volume shown in Fig. 2a. The base of the rectangular prism that we integrate over is given by the rectangle *RQMN*. The temporal extent of the rectangular prism is the time *T*. Using Gauss' law, volumetric integrations can be converted into surface integrals. The key difference between the integration performed here and that in Balsara [7] is that we also account for the integration over the resolved states stemming from the one-dimensional Riemann solvers, which yields a modicum of greater stability. Conceptually, both Balsara [7] and this paper use the same control volume from Fig. 2a to obtain the strongly-interacting state $\mathbf{U}^*$. The difference stems from the fact that here we choose *x-t* and *y-t* surfaces that are just inside the control volume shown in Fig. 2a with the result that we pick up the contribution from the resolved states of the one-dimensional Riemann problems. This is a stabilizing contribution. In Balsara [7] we use *x-t* and *y-t* surfaces that are just outside the same control volume, with the result that we do not get the contributions from the one-dimensional Riemann problems in those faces. Carrying out the above integration over the rectangular prism in Fig. 2a, along with a little simplification, allows us to write

$$\mathbf{U}^* = \frac{\mathbf{U}_{RU} S_R S_U + \mathbf{U}_{LD} S_L S_D - \mathbf{U}_{RD} S_R S_D - \mathbf{U}_{LU} S_L S_U}{(S_R - S_L)(S_U - S_D)}$$

$$- \frac{(\mathbf{F}_{RU} - \mathbf{F}_{LU}) S_U - (\mathbf{F}_{RD} - \mathbf{F}_{LD}) S_D + (\mathbf{G}_{RU} - \mathbf{G}_{RD}) S_R - (\mathbf{G}_{LU} - \mathbf{G}_{LD}) S_L}{(S_R - S_L)(S_U - S_D)}$$

$$+ \frac{(\mathbf{F}_{RU} - \mathbf{F}_R^*) S_U^R - (\mathbf{F}_{RD} - \mathbf{F}_R^*) S_D^R - (\mathbf{F}_{LU} - \mathbf{F}_L^*) S_U^L + (\mathbf{F}_{LD} - \mathbf{F}_L^*) S_D^L}{2 (S_R - S_L)(S_U - S_D)}$$

$$+ \frac{(\mathbf{G}_{RU} - \mathbf{G}_U^*) S_R^U - (\mathbf{G}_{LU} - \mathbf{G}_U^*) S_L^U - (\mathbf{G}_{RD} - \mathbf{G}_D^*) S_R^D + (\mathbf{G}_{LD} - \mathbf{G}_D^*) S_L^D}{2 (S_R - S_L)(S_U - S_D)}$$

(8)

When the variation is confined to the *x*-direction, the above expression successfully reduces to the resolved state for the one-dimensional HLL Riemann solver in that direction. I.e., the last two terms in the above equation become zero and the first two terms collapse to give the one-dimensional resolved state. Likewise, we get the desired simplifications for variations that are confined to the *y*-direction.

Obtaining the $x$-flux, $\mathbf{F}^*$, from the strongly-interacting region requires an integration of the conservation law over the $x \geq 0$ part of the volume shown in Fig. 2a. The base of the rectangular prism that we integrate over is given by the rectangle *DQMC*. The temporal extent of the rectangular prism is again given by the time *T*. Using Gauss' law, volumetric integrations can be converted into surface integrals. In order to make the integration easy and automatic, we define

$$S_R^{U+} = \max\left(S_R^U, 0\right) \;;\; S_L^{U+} = \max\left(S_L^U, 0\right) \;;\; S_R^{D+} = \max\left(S_R^D, 0\right) \;;\; S_L^{D+} = \max\left(S_L^D, 0\right) \tag{9}$$

Notice that a portion of the $x = 0$ plane is not swept over by the strongly-interacting state $\mathbf{U}^*$, as shown in Fig. 3a. Carrying out the integration gives us the expression for the *numerical x-flux* from the strongly-interacting region as

$$\begin{aligned}
\mathbf{F}^* = &\; 2\, \mathbf{U}^*\, S_R - \frac{S_U}{S_U - S_D}\mathbf{F}_U^{\text{HLL}} + \frac{S_D}{S_U - S_D}\mathbf{F}_D^{\text{HLL}} \\
&+ 2\,\frac{\mathbf{F}_{RU}\, S_U - \mathbf{F}_{RD}\, S_D + \left(\mathbf{G}_{RU} - \mathbf{G}_{RD}\right)S_R - \left(\mathbf{U}_{RU}\, S_U - \mathbf{U}_{RD}\, S_D\right)S_R}{S_U - S_D} \\
&+ \frac{\left(\mathbf{F}_{RD} - \mathbf{F}_R^*\right)S_D^R - \left(\mathbf{F}_{RU} - \mathbf{F}_R^*\right)S_U^R - \left(\mathbf{G}_{RU} - \mathbf{G}_U^*\right)S_R^{U+} + \left(\mathbf{G}_{LU} - \mathbf{G}_U^*\right)S_L^{U+} + \left(\mathbf{G}_{RD} - \mathbf{G}_D^*\right)S_R^{D+} - \left(\mathbf{G}_{LD} - \mathbf{G}_D^*\right)S_L^{D+}}{S_U - S_D}
\end{aligned}$$

(10)

The above $x$-flux is to be used when the strongly-interacting state straddles the time axis; i.e. when $S_L \leq 0 \leq S_R$ and $S_D \leq 0 \leq S_U$.

When the flow is strongly supersonic in both directions, we use the numerical $x$-flux from the appropriately upwinded quadrant in Fig. 1a. To make that explicit, we have

$$\begin{aligned}
\mathbf{F}^* =&\; \mathbf{F}_{LD} \text{ when } S_L \geq 0 \text{ and } S_D \geq 0 \\
=&\; \mathbf{F}_{LU} \text{ when } S_L \geq 0 \text{ and } S_U \leq 0 \\
=&\; \mathbf{F}_{RD} \text{ when } S_R \leq 0 \text{ and } S_D \geq 0 \\
=&\; \mathbf{F}_{RU} \text{ when } S_R \leq 0 \text{ and } S_U \leq 0
\end{aligned} \tag{11a}$$

We now consider the situation where we have supersonic flow in the $x$-direction, i.e. $S_L \geq 0$ or $S_R \leq 0$, along with sub-sonic flow in the $y$-direction, i.e. $S_D \leq 0 \leq S_U$. In this case we have some latitude in how the numerical $x$-fluxes are chosen. In the limit where $S_D \to 0_-$ we should have $\mathbf{F}^* \to \mathbf{F}_{LD}$ when $S_L \geq 0$ and $\mathbf{F}^* \to \mathbf{F}_{RD}$ when $S_R \leq 0$. In the limit where $S_U \to 0_+$ we should have $\mathbf{F}^* \to \mathbf{F}_{LU}$ when $S_L \geq 0$ and $\mathbf{F}^* \to \mathbf{F}_{RU}$ when $S_R \leq 0$. Recall that the results of the one-dimensional HLL Riemann solvers are available to us. So, all the limits in this discussion are cleanly accommodated by the formulae

$$\mathbf{F}^* = \frac{S_U}{S_U - S_D} \mathbf{F}_D^{HLL} - \frac{S_D}{S_U - S_D} \mathbf{F}_U^{HLL}$$

$$\mathbf{G}^* = \mathbf{G}_L^{HLL} \text{ when } S_L \geq 0 \tag{11b}$$
$$\quad\ \ = \mathbf{G}_R^{HLL} \text{ when } S_R \leq 0$$

This completes our derivation of the *x*-flux for the two-dimensional HLL Riemann solver.

Obtaining the *y*-flux, $\mathbf{G}^*$, from the strongly-interacting region requires an integration of the conservation law over the $y \geq 0$ part of the volume shown in Fig. 2a. The base of the rectangular prism that we integrate over is given by the rectangle *BAMN*. The time *T* gives the temporal extent of the rectangular prism. In order to make the integration easy and automatic, we define

$$S_U^{R+} = \max\left(S_U^R, 0\right) \ ; \ S_D^{R+} = \max\left(S_D^R, 0\right) \ ; \ S_U^{L+} = \max\left(S_U^L, 0\right) \ ; \ S_D^{L+} = \max\left(S_D^L, 0\right) \tag{12}$$

Notice that a portion of the $y = 0$ plane is not swept over by the strongly-interacting state $\mathbf{U}^*$, as shown in Fig. 3a. Carrying out the integration gives us the expression for the *numerical y-flux* from the strongly-interacting region as

$$\mathbf{G}^* = 2\,\mathbf{U}^* S_U - \frac{S_R}{S_R - S_L} \mathbf{G}_R^{HLL} + \frac{S_L}{S_R - S_L} \mathbf{G}_L^{HLL}$$
$$+ 2\,\frac{\mathbf{G}_{RU} S_R - \mathbf{G}_{LU} S_L + \left(\mathbf{F}_{RU} - \mathbf{F}_{LU}\right) S_U - \left(\mathbf{U}_{RU} S_R - \mathbf{U}_{LU} S_L\right) S_U}{S_R - S_L}$$
$$+ \frac{\left(\mathbf{G}_{LU} - \mathbf{G}_U^*\right) S_L^U - \left(\mathbf{G}_{RU} - \mathbf{G}_U^*\right) S_R^U - \left(\mathbf{F}_{RU} - \mathbf{F}_R^*\right) S_U^{R+} + \left(\mathbf{F}_{RD} - \mathbf{F}_R^*\right) S_D^{R+} + \left(\mathbf{F}_{LU} - \mathbf{F}_L^*\right) S_U^{L+} - \left(\mathbf{F}_{LD} - \mathbf{F}_L^*\right) S_D^{L+}}{S_R - S_L}$$

$$\tag{13}$$

The above *y*-flux is to be used when the strongly-interacting state straddles the time axis; i.e. when $S_L \leq 0 \leq S_R$ and $S_D \leq 0 \leq S_U$.

When the flow is strongly supersonic in both directions, we use the numerical *y*-flux from the appropriately upwinded quadrant in Fig. 1a. To make that explicit, we have

$$\mathbf{G}^* = \mathbf{G}_{LD} \text{ when } S_L \geq 0 \text{ and } S_D \geq 0$$
$$\quad\ \ = \mathbf{G}_{LU} \text{ when } S_L \geq 0 \text{ and } S_U \leq 0$$
$$\quad\ \ = \mathbf{G}_{RD} \text{ when } S_R \leq 0 \text{ and } S_D \geq 0 \tag{14a}$$
$$\quad\ \ = \mathbf{G}_{RU} \text{ when } S_R \leq 0 \text{ and } S_U \leq 0$$

We now consider the situation where we have supersonic flow in the *y*-direction, i.e. $S_D \geq 0$ or $S_U \leq 0$, along with sub-sonic flow in the *x*-direction, i.e. $S_L \leq 0 \leq S_R$. As before, we have some latitude in how the numerical *y*-fluxes are chosen. In the limit where $S_L \to 0_-$ we should have $\mathbf{G}^* \to \mathbf{G}_{LD}$ when $S_D \geq 0$ and $\mathbf{G}^* \to \mathbf{G}_{LU}$ when $S_U \leq 0$. In the limit where $S_R \to 0_+$ we should have $\mathbf{G}^* \to \mathbf{G}_{RD}$ when $S_D \geq 0$ and $\mathbf{G}^* \to \mathbf{G}_{RU}$ when $S_U \leq 0$. So, all the limits in this discussion are cleanly accommodated by the formulae

$$\mathbf{G}^* = \frac{S_R}{S_R - S_L}\mathbf{G}_L^{HLL} - \frac{S_L}{S_R - S_L}\mathbf{G}_R^{HLL}$$
$$\mathbf{F}^* = \mathbf{F}_D^{HLL} \text{ when } S_D \geq 0 \quad (14b)$$
$$\phantom{\mathbf{F}^*} = \mathbf{F}_U^{HLL} \text{ when } S_U \leq 0$$

This completes our derivation of the *y*-flux for the two-dimensional HLL Riemann solver.

There is indeed a simplifying assumption under which we can obtain very compact expressions for the strongly-interacting state and the corresponding numerical fluxes. It consists of setting $S_R^U$ and $S_R^D$ to $S_R$, the maximal right-going speed, and similarly setting $S_L^U$ and $S_L^D$ to $S_L$, the maximal left-going speed. In a similar fashion, we set $S_U^R$ and $S_U^L$ to $S_U$; and $S_D^R$ and $S_D^L$ to $S_D$. The strongly-interacting state is given by

$$\mathbf{U}^* = \frac{\mathbf{U}_{RU} S_R S_U + \mathbf{U}_{LD} S_L S_D - \mathbf{U}_{RD} S_R S_D - \mathbf{U}_{LU} S_L S_U}{(S_R - S_L)(S_U - S_D)}$$
$$- \frac{(\mathbf{F}_{RU} - \mathbf{F}_{LU})S_U - (\mathbf{F}_{RD} - \mathbf{F}_{LD})S_D + (\mathbf{G}_{RU} - \mathbf{G}_{RD})S_R - (\mathbf{G}_{LU} - \mathbf{G}_{LD})S_L}{2(S_R - S_L)(S_U - S_D)}$$
$$- \frac{\mathbf{F}_R^* - \mathbf{F}_L^*}{2(S_R - S_L)} - \frac{\mathbf{G}_U^* - \mathbf{G}_D^*}{2(S_U - S_D)} \quad (15)$$

We get a closed form expression for the numerical *x*-flux from the strongly-interacting state as

$$\mathbf{F}^* = \left[\frac{S_U}{S_U - S_D}\right]\mathbf{F}_U^{HLL} - \left[\frac{S_D}{S_U - S_D}\right]\mathbf{F}_D^{HLL} - \left[\frac{S_R S_L}{(S_R - S_L)(S_U - S_D)}\right](\mathbf{G}_{RU} - \mathbf{G}_{LU} + \mathbf{G}_{LD} - \mathbf{G}_{RD})$$
$$+ \frac{(\mathbf{F}_{RU} - \mathbf{F}_R^*) S_L S_U - (\mathbf{F}_{LU} - \mathbf{F}_L^*) S_R S_U + (\mathbf{F}_{LD} - \mathbf{F}_L^*) S_R S_D - (\mathbf{F}_{RD} - \mathbf{F}_R^*) S_L S_D}{(S_R - S_L)(S_U - S_D)}$$

(16)

The numerical *y*-flux from the strongly-interacting state yields

$$\mathbf{G}^* = \left[\frac{S_R}{S_R - S_L}\right]\mathbf{G}_R^{HLL} - \left[\frac{S_L}{S_R - S_L}\right]\mathbf{G}_L^{HLL} - \left[\frac{S_U S_D}{(S_R - S_L)(S_U - S_D)}\right](\mathbf{F}_{RU} - \mathbf{F}_{LU} + \mathbf{F}_{LD} - \mathbf{F}_{RD})$$
$$+ \frac{(\mathbf{G}_{RU} - \mathbf{G}_U^*) S_R S_D - (\mathbf{G}_{LU} - \mathbf{G}_U^*) S_L S_D + (\mathbf{G}_{LD} - \mathbf{G}_D^*) S_L S_U - (\mathbf{G}_{RD} - \mathbf{G}_D^*) S_R S_U}{(S_R - S_L)(S_U - S_D)}$$

(17)

The numerical fluxes in the supersonic cases are still given by eqns. (11) and (14). Please note that eqns. (15) to (17) are still HLL fluxes, not LLF fluxes (Rusanov [44]), because we do not set $S_R = -S_L \rightarrow \max(|S_R|, |S_L|)$ and $S_U = -S_D \rightarrow \max(|S_U|, |S_D|)$. If we do make that simplification, the above equations do collapse to their multidimensional LLF forms. This completes our description of the two-dimensional HLL Riemann solver with maximal speeds.

Several points are worth making about the results derived in this sub-section:

1) A numerical method that is built exclusively on the multi-dimensional Riemann solver technology presented here should be able to take larger time steps than equivalent schemes that are based on one-dimensional Riemann solver technology. We present such a scheme in a later section. A robust scheme that can take large time steps is very useful on parallel supercomputers because it means that the processors have to exchange fewer messages, resulting in better scalability.

2) Eqns. (15) through (17) have a compact and symmetrical form that is very well-suited for numerical implementation.

3) The compact form of eqns. (15) through (17) might make them useful when designing positivity proofs for the two-dimensional Riemann solvers presented here.

4) With the wave speeds frozen, eqns. (15) through (17) are linear in the four incoming states and their corresponding fluxes. This should make the present Riemann solver very useful for implicit treatment of hyperbolic systems where the hardest challenge is proper multi-dimensional coupling on the smallest possible stencil.

6) The full multi-dimensional HLL Riemann solver, as provided by eqns. (8) through (14), is also linear in the incoming states and their fluxes when the wave speeds are frozen. With a little algebra, its linear structure can be made as obviously transparent as eqns. (15) through (17).

7) With the wave speeds frozen, the Riemann solver's linear dependence on incoming states and their fluxes can also be used to advantage for several types of Hamilton-Jacobi problems.

8) Eqns. (15) through (17) do, however, have the disadvantage that they maximize the dissipation, but that may not be a big disadvantage in regions of smooth flow when higher order reconstruction is used. In regions where the flow has large discontinuities, the somewhat larger dissipation might indeed be an asset. If lower dissipation is desired, one should use eqns. (8) through (14). We have used eqns. (8) through (14) (and their HLLC extensions) exclusively in all our test problems and found them to perform very robustly in the vicinity of ultra-strong shocks.

**II.b) Derivation of the HLLC Riemann Solver in Two Dimensions**

Now that we have a complete derivation of the two-dimensional HLL Riemann solver, we embark on a derivation of the two-dimensional HLLC Riemann solver. From Schulz-Rinne,

Collins and Glaz [47] we see that contact discontinuities that are oriented in any possible direction on the computational mesh constitute one of the persistent flow structures in the strong-interaction region of many multi-dimensional Riemann problems. For that reason, we want our multi-dimensional HLLC Riemann solver to resolve contact discontinuities in the region of strong interaction. Figs. 1b, 2b and 3b depict the changes in the wave model when an intermediate wave, corresponding to the contact discontinuity, is introduced in the original strongly-interacting state. At each stage in the discussion, these figures should be compared to Figs. 1a, 2a and 3a to appreciate the differences between the multidimensional HLL and HLLC Riemann solvers. The original strongly-interacting state is now sub-divided into two strongly-interacting states that are separated by a contact discontinuity. In keeping with the spirit of a multidimensional approach, the discontinuity can move in any direction in the *x-y* plane with *x*- and *y*-speeds given by $S_{Mx}^*$ and $S_{My}^*$. The orientation of the contact discontinuity is set by an examination of the density gradient in the larger scale flow. The velocity with which the contact discontinuity propagates, as well as the extent of the density jump across it, are set by the Riemann solver.

We define the unit vectors along the *x*-, *y*- and *t*-directions by $\mathbf{i}$, $\mathbf{j}$ and $\mathbf{t}$. The configuration depicted in Figs. 1b and 3b is self-similar. Thus, like the one-dimensional Riemann problem, our two-dimensional Riemann problem has the essential property of being self-similar. In Fig. 1b, the contact discontinuity separates the two strongly-interacting states given by $\mathbf{U}_{C1}^*$ and $\mathbf{U}_{C2}^*$. We establish the convention that $\mathbf{U}_{C1}^*$ is the state that covers the origin *O* and is, therefore, the state of greater computational interest. The physics of contact discontinuities tells us that a discontinuity can be passively advected by the velocity vector. As a result, the contact discontinuity can make any angle to the velocity vector and we hope that Fig. 1b also illustrates this. The angle that the contact discontinuity makes relative to the mesh can only be determined by evaluating the gradient of the density from the four zones that come together at the edge. As a result, we assume that the normal vector to the contact discontinuity is also provided to the Riemann solver.

Fig. 2b is analogous to Fig. 2a and we use it to define many of the fluxes and states that are useful in the construction of the multi-dimensional HLLC Riemann solver. We now have an *x*-directional HLLC Riemann solver between the states $\mathbf{U}_{LD}$ and $\mathbf{U}_{RD}$ in the lower face. The maximal speeds $S_R^D$ and $S_L^D$ are still present in the Riemann solver. We also have a contact speed of $S_M^D$ separating two constant resolved states $\mathbf{U}_D^{*-}$ and $\mathbf{U}_D^{*+}$, and a numerical *x*-flux $\mathbf{F}_D^{HLLC}$. Corresponding to these two resolved states, we can obtain *y*-fluxes given by $\mathbf{G}_D^{*-}$ and $\mathbf{G}_D^{*+}$ respectively. Similarly, the right face has a *y*-directional HLLC Riemann solver between the states $\mathbf{U}_{RD}$ and $\mathbf{U}_{RU}$ with maximal speeds $S_U^R$ and $S_D^R$. We also have a contact speed of $S_M^R$ separating two constant resolved states $\mathbf{U}_R^{*-}$ and $\mathbf{U}_R^{*+}$, and a numerical *y*-flux $\mathbf{G}_R^{HLLC}$. Corresponding to these two resolved states, we have *x*-fluxes given by $\mathbf{F}_R^{*-}$ and $\mathbf{F}_R^{*+}$ respectively. The upper face is hidden from view in Fig. 2b, but it has an *x*-directional HLLC Riemann solver between the states $\mathbf{U}_{LU}$ and $\mathbf{U}_{RU}$ with maximal speeds $S_R^U$ and $S_L^U$. Additionally, we have a contact speed of $S_M^U$ separating two constant resolved states $\mathbf{U}_U^{*-}$ and $\mathbf{U}_U^{*+}$, and a numerical *x*- flux $\mathbf{F}_U^{HLLC}$. Corresponding to these two resolved states, we have *y*-fluxes given by $\mathbf{G}_U^{*-}$ and $\mathbf{G}_U^{*+}$ respectively. Similarly, the left face is hidden from view in Fig. 2b, however it has a *y*-directional HLLC Riemann solver between the states $\mathbf{U}_{LD}$ and $\mathbf{U}_{LU}$ with maximal speeds $S_U^L$ and $S_D^L$. This yields a contact speed of $S_M^L$ separating two constant resolved states $\mathbf{U}_L^{*-}$ and $\mathbf{U}_L^{*+}$, and a numerical *y*-flux $\mathbf{G}_L^{HLLC}$. Corresponding to these two resolved states, we have *x*-fluxes given by $\mathbf{F}_L^{*-}$ and $\mathbf{F}_L^{*+}$ respectively. These *x*- and *y*-directional HLLC Riemann solvers help reduce the dissipation in the two-dimensional HLLC Riemann solver relative to the two-dimensional HLL Riemann solver. For much of the rest of this sub-section, we will focus on flows that are subsonic in both directions so that the strongly-interacting state straddles the time axis.

The final, two-dimensional, strongly-interacting states, $\mathbf{U}_{C1}^*$ and $\mathbf{U}_{C2}^*$, are shown in Fig. 3b. That figure also shows the space-time volume occupied by these states. Analogously to Fig. 3a, this volume is an inverted rectangular pyramid. We will use sub-portions of this inverted rectangular pyramid as our control volume in space and time. Specifically, we will focus on the portion of the pyramid shown in Fig. 3b that contains the state $\mathbf{U}_{C1}^*$. The strongly-interacting state $\mathbf{U}_{C1}^*$ is bounded by the thickened lines of Fig. 3b which show the portions of the faces over which we will have to integrate area integrals. Because Fig. 3b shows a three-dimensional shape, we can only show the nearest two triangular faces of the inverted pyramid, along with its rectangular base, over which we might have to evaluate area integrals. The sub-portion of the inverted pyramid that is bounded by the thick lines in Fig. 3b forms the three-dimensional control volume in space-time over which we will integrate to obtain the strongly-interacting state $\mathbf{U}_{C1}^*$. The reason for picking such a control volume now becomes evident. All the surface integrals on this control volume will reduce to integrating over triangular or trapezoidal shapes with areas and outward facing normals that are easily evaluated. Since triangles and trapezoids are amongst the simplest geometric shapes, our reason for picking the wave model that was adopted in this paper now becomes self-evident – it yields the simplest strategy for evaluating the space-time integrals that are needed in our multi-dimensional Riemann solver.

Every HLLC Riemann solver is built on top of an HLL Riemann solver and the present one is no exception. Thus we first use eqns. (8), (10) and (13) to obtain $\mathbf{U}^*$, $\mathbf{F}^*$, and $\mathbf{G}^*$. Now that the mean, strongly-interacting state $\mathbf{U}^*$ for the two-dimensional HLL Riemann solver is available, we concern ourselves with the definition of the *x*- and *y*-speeds of the contact discontinuity, given by $S_{Mx}^*$ and $S_{My}^*$. We, therefore, follow the suggestion of Toro, Spruce & Speares [48] and Batten *et al.* [11] and define

$$\rho^* = \left(\mathbf{U}^*\right)_1 \quad ; \quad S_{Mx}^* = \frac{\left(\mathbf{U}^*\right)_2}{\rho^*} \quad ; \quad S_{My}^* = \frac{\left(\mathbf{U}^*\right)_3}{\rho^*} \tag{18}$$

where the subscripts in the lower right corners of the above brackets denote their corresponding components. Here $\rho^*$ has units of density. The *x*-velocity of the contact, $S_{Mx}^*$, is given by the ratio of the second component of the strongly-interacting state $\mathbf{U}^*$ to its first component;

analogously for $S_{My}^*$. The velocity defined by the above equation is taken to be the same for both the constant states $\mathbf{U}_{C1}^*$ and $\mathbf{U}_{C2}^*$. Now we need to operationally definine the pressure in the strongly-interacting state. Again, following the cue that has been established, we can define the pressure in the strongly-interacting state for the Euler system as

$$P^* = \frac{1}{2}\left\{\left[\left(\mathbf{F}^*\right)_2 - \rho^*\left(S_{Mx}^*\right)^2\right] + \left[\left(\mathbf{G}^*\right)_3 - \rho^*\left(S_{My}^*\right)^2\right]\right\} \tag{19}$$

The pressure defined by the above equation is taken to be the same for both the constant states $\mathbf{U}_{C1}^*$ and $\mathbf{U}_{C2}^*$. Now that $P^*$, $S_{Mx}^*$ and $S_{My}^*$ are available for $\mathbf{U}^*$, we can formally write down the expressions for the *numerical x*- and *y*-fluxes in the strongly-interacting state $\mathbf{U}_{C1}^*$ of interest as

$$\mathbf{F}_{C1}^* = S_{Mx}^* \mathbf{U}_{C1}^* + \left(0, P^*, 0, 0, P^* S_{Mx}^*\right)^T \quad ; \quad \mathbf{G}_{C1}^* = S_{My}^* \mathbf{U}_{C1}^* + \left(0, 0, P^*, 0, P^* S_{My}^*\right)^T \tag{20}$$

In other words, the fluxes in the *x*- and *y*-directions are linear combinations of $\mathbf{U}_{C1}^*$ along with the addition of a constant vector that is already known. The above fluxes would be completely specified if $\mathbf{U}_{C1}^*$ were specified. For that reason, from the next paragraph onwards, we embark on a strategy for obtaining $\mathbf{U}_{C1}^*$.

Our task in the rest of this sub-section is to find an implementable strategy for evaluating $\mathbf{U}_{C1}^*$. For that reason, the description of the rest of this sub-section is rather operational and resembles pseudo-code. Without the aid of such a description, the reader would find it very difficult to implement the two-dimensional HLLC Riemann solver. Let the velocity vector in the *x-y* plane for both the strongly-interacting states be given by $\mathbf{S}_M^* \equiv S_{Mx}^* \mathbf{i} + S_{My}^* \mathbf{j}$, and let the contact discontinuity have a unit normal that is given by $\boldsymbol{\sigma} \equiv \sigma_x \mathbf{i} + \sigma_y \mathbf{j}$. For full generality, the velocity vector of the strongly-interacting states does not have to be orthogonal to the contact discontinuity, i.e. vectors $\mathbf{S}_M^*$ and $\boldsymbol{\sigma}$ are not parallel. The locus of the contact discontinuity at time $T$ in Fig. 1b is given by

$$y = S_{My}^* T - \frac{\sigma_x}{\sigma_y}\left(x - S_{Mx}^* T\right) \tag{21}$$

To efficiently evaluate the facial contributions to $\mathbf{U}_{C1}^*$ from the various faces of the pyramid shown in Fig. 3b, we first need to evaluate the intersections of the contact discontinuity with the various faces of the rectangle shown in Fig. 4. The rectangle in Fig. 4 shows a projection of the rectangular base of the inverted pyramid from Fig. 3b on to the x-y plane. The contact discontinuity intersects $QM$ at $\left(S_R T, Y_R T, T\right)$ and $RN$ at $\left(S_L T, Y_L T, T\right)$, where the coordinates refer to the *x*-, *y*- and *t*-coordinates in a three-dimensional space-time. Similarly, it intersects $MN$ at $\left(X_U T, S_U T, T\right)$ and $QR$ at $\left(X_D T, S_D T, T\right)$. The previous two sentences serve to define $Y_R$, $Y_L$, $X_U$ and $X_D$, which are given by

$$Y_R = S^*_{My} - \sigma_x \left(S_R - S^*_{Mx}\right)/\sigma_y \quad ; \quad Y_L = S^*_{My} - \sigma_x \left(S_L - S^*_{Mx}\right)/\sigma_y \quad ;$$
$$X_U = S^*_{Mx} - \sigma_y \left(S_U - S^*_{My}\right)/\sigma_x \quad ; \quad X_D = S^*_{Mx} - \sigma_y \left(S_D - S^*_{My}\right)/\sigma_x \quad \quad (22)$$

The above equation helps us in finding the area occupied by $\mathbf{U}^*_{C1}$ in the base of the inverted pyramid. We describe an automated strategy for evaluating this area in the next paragraph.

The thick boundary in Fig. 4 highlights the part of the base of the pyramid at time $T$ that contains the strongly-interacting state $\mathbf{U}^*_{C1}$. We will be integrating over this area, so let it be denoted by $A^*_{C1}T^2$. We can now obtain a closed form expression for $A^*_{C1}$ pertaining to the six different situations that prevail as follows if $\left(S_L \leq X_D \leq S_R \text{ and } S_L \leq X_U \leq S_R\right)$ then

$$A_{C1}^* = \frac{1}{2}(X_U + X_D - 2S_L)(S_U - S_D) \quad \text{when } X_D S_U - S_D X_U > 0$$

$$A_{C1}^* = \frac{1}{2}(2S_R - X_U - X_D)(S_U - S_D) \quad \text{otherwise}$$

else if $\left(S_D \leq Y_L \leq S_U \text{ and } S_D \leq Y_R \leq S_U\right)$ then

$$A_{C1}^* = \frac{1}{2}(S_R - S_L)(Y_R + Y_L - 2S_D) \quad \text{when } S_R Y_L - Y_R S_L > 0$$

$$A_{C1}^* = \frac{1}{2}(S_R - S_L)(2S_U - Y_R - Y_L) \quad \text{otherwise}$$

else if $\left(S_L < X_U < S_R \text{ and } S_D < Y_R < S_U\right)$ then

$$A_{C1}^* = (S_R - S_L)(S_U - S_D) - \frac{1}{2}(S_R - X_U)(S_U - Y_R) \quad \text{when } S_R S_U - Y_R X_U > 0$$

$$A_{C1}^* = \frac{1}{2}(S_R - X_U)(S_U - Y_R) \quad \text{otherwise}$$

else if $\left(S_L < X_D < S_R \text{ and } S_D < Y_L < S_U\right)$ then

$$A_{C1}^* = \frac{1}{2}(X_D - S_L)(Y_L - S_D) \quad \text{when } X_D Y_L - S_D S_L > 0$$

$$A_{C1}^* = (S_R - S_L)(S_U - S_D) - \frac{1}{2}(X_D - S_L)(Y_L - S_D) \quad \text{otherwise}$$

else if $\left(S_L < X_U < S_R \text{ and } S_D < Y_L < S_U\right)$ then

$$A_{C1}^* = (S_R - S_L)(S_U - S_D) - \frac{1}{2}(X_U - S_L)(S_U - Y_L) \quad \text{when } X_U Y_L - S_U S_L > 0$$

$$A_{C1}^* = \frac{1}{2}(X_U - S_L)(S_U - Y_L) \quad \text{otherwise}$$

$$(23)$$

else if $\left(S_L < X_D < S_R \text{ and } S_D < Y_R < S_U\right)$ then

$$A_{C1}^* = \frac{1}{2}(Y_R - S_D)(S_R - X_D) \quad \text{when } S_R S_D - Y_R X_D > 0$$

$$A_{C1}^* = (S_R - S_L)(S_U - S_D) - \frac{1}{2}(Y_R - S_D)(S_R - X_D) \quad \text{otherwise}$$

end if

For each of the above six orientations of the contact discontinuity identified above, the area of interest is identified as the one that contains the *t*-axis in Fig. 3b. This is also the area that contains the strongly-interacting state $\mathbf{U}_{C1}^*$ that is of interest to us. The corresponding numerical fluxes are computed with the help of the state $\mathbf{U}_{C1}^*$, see eqn. (20). The area integrals in the triangular faces of the inverted pyramid shown in Fig. 3b are rather easy to evaluate and will be addressed in due time.

The area integral that is hardest to evaluate is the one separating the two strongly-interacting states, $\mathbf{U}^*_{C1}$ and $\mathbf{U}^*_{C2}$. This area is given by the triangle *OJK* in Fig. 3b. We draw on the insight from physics that the cross product of two vectors yields a vector that is orthogonal to each of the two vectors and whose magnitude is equal to twice the area of the triangle contained between the two vectors. In physics, area can be viewed as a vector. As a result, the outward pointing area vector for the triangle that separates $\mathbf{U}^*_{C1}$ and $\mathbf{U}^*_{C2}$ is specified by

if $(S_L \leq X_D \leq S_R$ and $S_L \leq X_U \leq S_R)$ then

$$N_x^{C1C2}\mathbf{i} + N_y^{C1C2}\mathbf{j} + N_t^{C1C2}\mathbf{t} = \xi \frac{1}{2}(X_U\mathbf{i} + S_U\mathbf{j} + \mathbf{t}) \times (X_D\mathbf{i} + S_D\mathbf{j} + \mathbf{t}) \quad \text{with} \quad \xi \equiv \text{sgn}(X_D S_U - S_D X_U)$$

else if $(S_D \leq Y_L \leq S_U$ and $S_D \leq Y_R \leq S_U)$ then

$$N_x^{C1C2}\mathbf{i} + N_y^{C1C2}\mathbf{j} + N_t^{C1C2}\mathbf{t} = \xi \frac{1}{2}(S_L\mathbf{i} + Y_L\mathbf{j} + \mathbf{t}) \times (S_R\mathbf{i} + Y_R\mathbf{j} + \mathbf{t}) \quad \text{with} \quad \xi \equiv \text{sgn}(S_R Y_L - Y_R S_L)$$

else if $(S_L < X_U < S_R$ and $S_D < Y_R < S_U)$ then

$$N_x^{C1C2}\mathbf{i} + N_y^{C1C2}\mathbf{j} + N_t^{C1C2}\mathbf{t} = \xi \frac{1}{2}(X_U\mathbf{i} + S_U\mathbf{j} + \mathbf{t}) \times (S_R\mathbf{i} + Y_R\mathbf{j} + \mathbf{t}) \quad \text{with} \quad \xi \equiv \text{sgn}(S_R S_U - Y_R X_U)$$

else if $(S_L < X_D < S_R$ and $S_D < Y_L < S_U)$ then

$$N_x^{C1C2}\mathbf{i} + N_y^{C1C2}\mathbf{j} + N_t^{C1C2}\mathbf{t} = \xi \frac{1}{2}(S_L\mathbf{i} + Y_L\mathbf{j} + \mathbf{t}) \times (X_D\mathbf{i} + S_D\mathbf{j} + \mathbf{t}) \quad \text{with} \quad \xi \equiv \text{sgn}(X_D Y_L - S_D S_L)$$

else if $(S_L < X_U < S_R$ and $S_D < Y_L < S_U)$ then

$$N_x^{C1C2}\mathbf{i} + N_y^{C1C2}\mathbf{j} + N_t^{C1C2}\mathbf{t} = \xi \frac{1}{2}(S_L\mathbf{i} + Y_L\mathbf{j} + \mathbf{t}) \times (X_U\mathbf{i} + S_U\mathbf{j} + \mathbf{t}) \quad \text{with} \quad \xi \equiv \text{sgn}(X_U Y_L - S_U S_L)$$

else if $(S_L < X_D < S_R$ and $S_D < Y_R < S_U)$ then

$$N_x^{C1C2}\mathbf{i} + N_y^{C1C2}\mathbf{j} + N_t^{C1C2}\mathbf{t} = \xi \frac{1}{2}(X_D\mathbf{i} + S_D\mathbf{j} + \mathbf{t}) \times (S_R\mathbf{i} + Y_R\mathbf{j} + \mathbf{t}) \quad \text{with} \quad \xi \equiv \text{sgn}(S_R S_D - Y_R X_D)$$

end if

(24)

Eqn. (24) serves to define the components $N_x^{C1C2}$, $N_y^{C1C2}$ and $N_t^{C1C2}$ of the outward pointing area vector. Notice that the area vectors are obtained by applying the right hand rule to triangle *OJK*, with the points being traversed in that sequence. With these areas in hand, we can evaluate their dot products with $\mathbf{F}^*_{C1}\mathbf{i} + \mathbf{G}^*_{C1}\mathbf{j} + \mathbf{U}^*_{C1}\mathbf{t}$ in order to obtain the area integral over the triangle that separates the two strongly-interacting states.

We can now write an equation for $\mathbf{U}^*_{C1}$ after removing an obvious factor of $T^2$ as

$$\left(A_{C1}^* + N_t^{C1C2} + N_x^{C1C2} S_{Mx}^* + N_y^{C1C2} S_{My}^*\right)\mathbf{U}_{C1}^* = -N_x^{C1C2}\left(0, P^*, 0, 0, P^* S_{Mx}^*\right)^T - N_y^{C1C2}\left(0, 0, P^*, 0, P^* S_{My}^*\right)^T$$

$$-\text{(Sums over triangles in the upper face)}$$
$$-\text{(Sums over triangles in the lower face)}$$
$$-\text{(Sums over triangles in the right face)}$$
$$-\text{(Sums over triangles in the left face)}$$

(25)

Each of the last four sums are either evaluated over the whole relevant triangular face of the pyramid, or they are evaluated over a sub-portion of it. Each of those four sums is evaluated similarly, so our description would be complete if we demonstrate one of them. Since the lower triangular face of the inverted pyramid in Fig. 3b faces us, it is best to show how the sums are evaluated in that face. Fig. 5 shows us how the wave speeds and states are labeled and ordered on each of the faces. It also gives the names of the states that lie between the waves. It is very useful in identifying the limits of integration over the triangular faces of the inverted pyramid shown in Fig. 3b.

We first illustrate the integration over the lower triangular face for the situation shown in Fig. 3b where $S_L \leq X_D \leq S_R$ and $S_L \leq X_U \leq S_R$ and $X_D S_U - S_D X_U > 0$. The areal integration extends over the triangle with space-time vertices given by $(0,0,0)$, $(X_D T, S_D T, T)$ and $(S_L T, S_D T, T)$; where the right hand rule would produce an outward pointing normal if the three points were traversed in sequence. We refer to this as a *Type I integration over the lower triangular face*, and we point out that it corresponds to the range $x \in [S_L, X_D]$ of the lower boundary of Fig. 5. The task of evaluating the area integral reduces to a matter of figuring out where $X_D$ is located relative to the wave speeds at the lower face. We present pseudocode for carrying out the area integral over the lower triangular face of the inverted pyramid in Fig. 3b below. An obvious factor of $T^2$ has been removed from the area integral. The limits of integration over each sub-triangle that corresponds to a constant state are denoted $X_1$ and $X_2$ below, where we have $X_2 > X_1$.

$(\text{Sums over triangles in the lower face}) = 0$

$if\ (X_D > S_R^D)\{$

$\quad X_2 = \min(X_D, S_R)\ ;\ X_1 = S_R^D\ ;$

$\quad (\text{Sums over triangles in the lower face}) += \dfrac{(X_2 - X_1)}{2}(-\mathbf{G}_{RD} + S_D \mathbf{U}_{RD})\ ;$

$\}$

$if\ (X_D > S_M^D)\{$

$\quad X_2 = \min(X_D, S_R^D)\ ;\ X_1 = S_M^D\ ;$

$\quad (\text{Sums over triangles in the lower face}) += \dfrac{(X_2 - X_1)}{2}(-\mathbf{G}_D^{*+} + S_D \mathbf{U}_D^{*+})\ ;$

$\}$

$if\ (X_D > S_L^D)\{$

$\quad X_2 = \min(X_D, S_M^D)\ ;\ X_1 = S_L^D\ ;$

$\quad (\text{Sums over triangles in the lower face}) += \dfrac{(X_2 - X_1)}{2}(-\mathbf{G}_D^{*-} + S_D \mathbf{U}_D^{*-})\ ;$

$\}$

$if\ (X_D > S_L)\{$

$\quad X_2 = \min(X_D, S_L^D)\ ;\ X_1 = S_L\ ;$

$\quad (\text{Sums over triangles in the lower face}) += \dfrac{(X_2 - X_1)}{2}(-\mathbf{G}_{LD} + S_D \mathbf{U}_{LD})\ ;$

$\}$

(26)

This completes our description of Type I integration at the lower triangular face. There are three more types of integration that one may need in the lower triangular face and they are described in Appendix A. Integrals in the other three triangular faces are constructed similarly and it would be redundant to explicit all of them.

This is the all-important paragraph for this sub-section where we provide a pointwise description of how the numerical fluxes, $\mathbf{F}^*$ and $\mathbf{G}^*$, in the two-dimensional HLLC Riemann solver are assembled at the edges of the computational mesh:

1) We first consider the situation when the strongly-interacting state $\mathbf{U}_{C1}^*$ straddles the *t*-axis, i.e. when $S_L \leq 0 \leq S_R$ and $S_D \leq 0 \leq S_U$. Once the strongly-interacting state $\mathbf{U}_{C1}^*$ is obtained from eqn. (25), the associated numerical *x*- and *y*-fluxes are easily obtained from eqn. (20).

2) When the flow is supersonic in both directions, the upwinded numerical fluxes are trivially obtained from eqns. (11a) and (14a).

3) When the region of strong interaction is subsonic in the $y$-direction (i.e. $S_D \leq 0 \leq S_U$) and supersonic in the $x$-direction (i.e. $S_L \geq 0$ or $S_R \leq 0$), the numerical fluxes are given by

$$\mathbf{F}^* = \frac{S_U}{S_U - S_D} \mathbf{F}_D^{\text{HLLC}} - \frac{S_D}{S_U - S_D} \mathbf{F}_U^{\text{HLLC}}$$
$$\mathbf{G}^* = \mathbf{G}_L^{HLLC} \text{ when } S_L \geq 0 \qquad (27)$$
$$\phantom{\mathbf{G}^*} = \mathbf{G}_R^{HLLC} \text{ when } S_R \leq 0$$

4) When the region of strong interaction is subsonic in the $x$-direction (i.e. $S_L \leq 0 \leq S_R$) and supersonic in the $y$-direction (i.e. $S_D \geq 0$ or $S_U \leq 0$), the numerical fluxes are given by

$$\mathbf{G}^* = \frac{S_R}{S_R - S_L} \mathbf{G}_L^{HLLC} - \frac{S_L}{S_R - S_L} \mathbf{G}_R^{HLLC}$$
$$\mathbf{F}^* = \mathbf{F}_D^{HLLC} \text{ when } S_D \geq 0 \qquad (28)$$
$$\phantom{\mathbf{F}^*} = \mathbf{F}_U^{HLLC} \text{ when } S_U \leq 0$$

Eqns. (20), (25), (11a), (14a), (27) and (28) are useful for implementing the results of this section into computer code. They yield multidimensionally upwinded fluxes at the edges of the computational mesh. Sub-section IV.a provides further details on turning those fluxes into face-centered fluxes. This completes our description of the two-dimensional HLLC Riemann solver. In Appendix B we provide a simple strategy for obtaining the strongly interacting state $\mathbf{U}_{C2}^*$ once the strongly interacting state $\mathbf{U}_{C1}^*$ is known. This might be useful for making analyses of the Riemann solver.

### III) Modifications for the HLLC Riemann solver for MHD

To show that the method presented in the previous section is very general, we now present an HLLC formulation for MHD. The MHD equations can be written in two dimensions as a conservation law, $\partial_t \mathbf{U} + \partial_x \mathbf{F} + \partial_y \mathbf{G} = 0$, given by

$$\frac{\partial}{\partial t}\begin{pmatrix}\rho\\ \rho v_x\\ \rho v_y\\ \rho v_z\\ \varepsilon\\ B_x\\ B_y\\ B_z\end{pmatrix}+\frac{\partial}{\partial x}\begin{pmatrix}\rho v_x\\ \rho v_x^2+P+\mathbf{B}^2/8\pi-B_x^2/4\pi\\ \rho v_x v_y-B_x B_y/4\pi\\ \rho v_x v_z-B_x B_z/4\pi\\ (\varepsilon+P+\mathbf{B}^2/8\pi)v_x-B_x(\mathbf{v}\cdot\mathbf{B})/4\pi\\ 0\\ (v_x B_y-v_y B_x)\\ -(v_z B_x-v_x B_z)\end{pmatrix}+\frac{\partial}{\partial y}\begin{pmatrix}\rho v_y\\ \rho v_x v_y-B_x B_y/4\pi\\ \rho v_y^2+P+\mathbf{B}^2/8\pi-B_y^2/4\pi\\ \rho v_y v_z-B_y B_z/4\pi\\ (\varepsilon+P+\mathbf{B}^2/8\pi)v_y-B_y(\mathbf{v}\cdot\mathbf{B})/4\pi\\ -(v_x B_y-v_y B_x)\\ 0\\ (v_y B_z-v_z B_y)\end{pmatrix}=0$$

(29)

where $\rho$ is the density; $v_x$, $v_y$ and $v_z$ are the velocity components; $B_x$, $B_y$ and $B_z$ are the magnetic field components; $\varepsilon = \rho\,\mathbf{v}^2/2 + P/(\gamma-1) + \mathbf{B}^2/8\pi$ is the total energy and $\gamma$ is the ratio of specific heats. Notice that the *z*-component of the field can change in response to the other changes, even though the *z*-variation is suppressed. The crux of the formulation is that the flux vectors can be written as a linear combination of the conserved variables along with the addition of some extra terms. Using the strongly-interacting state $\mathbf{U}^*$ that comes from the two-dimensional HLL Riemann solver we can then write

$$\rho^* = (\mathbf{U}^*)_1 \;;\; S_{Mx}^* = \frac{(\mathbf{U}^*)_2}{\rho^*} \;;\; S_{My}^* = \frac{(\mathbf{U}^*)_3}{\rho^*} \;;\; S_{Mz}^* = \frac{(\mathbf{U}^*)_4}{\rho^*} \;;\; B_x^* = (\mathbf{U}^*)_6 \;;\; B_y^* = (\mathbf{U}^*)_7 \;;\; B_z^* = (\mathbf{U}^*)_8$$

(30)

The sum of the thermal and magnetic pressures, $P_M^*$, is given by the average of the total pressure obtained from the second component of the *x*-flux and the third component of the *y*-flux. We have

$$P_M^* = \frac{1}{2}\left\{\left[(\mathbf{F}^*)_2 - \rho^*\left(S_{Mx}^*\right)^2 + \left(B_x^*\right)^2/4\pi\right] + \left[(\mathbf{G}^*)_3 - \rho^*\left(S_{My}^*\right)^2 + \left(B_y^*\right)^2/4\pi\right]\right\}$$

(31)

Here $\mathbf{F}^*$ and $\mathbf{G}^*$ are obtained from the two-dimensional HLL Riemann solver. The *x*- and *y*-fluxes for the HLLC Riemann solver can now be written as

$$\mathbf{F}_{C1}^{*} = S_{Mx}^{*}\, \mathbf{U}_{C1}^{*} + \begin{pmatrix} 0 \\ P_{M}^{*} - \left(B_{x}^{*}\right)^{2}/4\pi \\ -B_{x}^{*}B_{y}^{*}/4\pi \\ -B_{x}^{*}B_{z}^{*}/4\pi \\ P_{M}^{*}S_{Mx}^{*} - B_{x}^{*}\left(\mathbf{S}^{*}\bullet\mathbf{B}^{*}\right)/4\pi \\ -S_{Mx}^{*}B_{x}^{*} \\ -S_{My}^{*}B_{x}^{*} \\ -S_{Mz}^{*}B_{x}^{*} \end{pmatrix} \quad ; \quad \mathbf{G}_{C1}^{*} = S_{My}^{*}\, \mathbf{U}_{C1}^{*} + \begin{pmatrix} 0 \\ -B_{x}^{*}B_{y}^{*}/4\pi \\ P_{M}^{*} - \left(B_{y}^{*}\right)^{2}/4\pi \\ -B_{y}^{*}B_{z}^{*}/4\pi \\ P_{M}^{*}S_{My}^{*} - B_{y}^{*}\left(\mathbf{S}^{*}\bullet\mathbf{B}^{*}\right)/4\pi \\ -S_{Mx}^{*}B_{y}^{*} \\ -S_{My}^{*}B_{y}^{*} \\ -S_{Mz}^{*}B_{y}^{*} \end{pmatrix}$$

(32)

We can now write an equation for $\mathbf{U}_{C1}^{*}$ as

$$\left(A_{C1}^{*} + N_{t}^{C1C2} + N_{x}^{C1C2}S_{Mx}^{*} + N_{y}^{C1C2}S_{My}^{*}\right)\mathbf{U}_{C1}^{*} = -N_{x}^{C1C2}\begin{pmatrix} 0 \\ P_{M}^{*} - \left(B_{x}^{*}\right)^{2}/4\pi \\ -B_{x}^{*}B_{y}^{*}/4\pi \\ -B_{x}^{*}B_{z}^{*}/4\pi \\ P_{M}^{*}S_{Mx}^{*} - B_{x}^{*}\left(\mathbf{S}^{*}\bullet\mathbf{B}^{*}\right)/4\pi \\ -S_{Mx}^{*}B_{x}^{*} \\ -S_{My}^{*}B_{x}^{*} \\ -S_{Mz}^{*}B_{x}^{*} \end{pmatrix} - N_{y}^{C1C2}\begin{pmatrix} 0 \\ -B_{x}^{*}B_{y}^{*}/4\pi \\ P_{M}^{*} - \left(B_{y}^{*}\right)^{2}/4\pi \\ -B_{y}^{*}B_{z}^{*}/4\pi \\ P_{M}^{*}S_{My}^{*} - B_{y}^{*}\left(\mathbf{S}^{*}\bullet\mathbf{B}^{*}\right)/4\pi \\ -S_{Mx}^{*}B_{y}^{*} \\ -S_{My}^{*}B_{y}^{*} \\ -S_{Mz}^{*}B_{y}^{*} \end{pmatrix}$$

$$-\left(\text{Sums over triangles in the upper face}\right) - \left(\text{Sums over triangles in the lower face}\right)$$
$$-\left(\text{Sums over triangles in the right face}\right) - \left(\text{Sums over triangles in the left face}\right)$$

(33)

As pointed out by Li [35], the one-dimensional HLLC Riemann solver for MHD suffers a slight loss of consistency in the magnetic field variables when the flux is written in term of the strongly-interacting state in eqn. (32). The suggestion for restoring consistency is to replace the magnetic field in the strongly-interacting state from the previous equation by the magnetic field from the HLL flux. We follow this suggestion in our two-dimensional HLLC Riemann solver. Thus, after $\mathbf{U}_{C1}^{*}$ is obtained and $\mathbf{F}_{C1}^{*}$ and $\mathbf{G}_{C1}^{*}$ are evaluated, we reset the sixth, seventh and eighth components as follows

$$\left(\mathbf{U}_{C1}^{*}\right)_{6} = \left(\mathbf{U}^{*}\right)_{6} \; ; \; \left(\mathbf{U}_{C1}^{*}\right)_{7} = \left(\mathbf{U}^{*}\right)_{7} \; ; \; \left(\mathbf{U}_{C1}^{*}\right)_{8} = \left(\mathbf{U}^{*}\right)_{8} \; ;$$
$$\left(\mathbf{F}_{C1}^{*}\right)_{6} = \left(\mathbf{F}^{*}\right)_{6} \; ; \; \left(\mathbf{F}_{C1}^{*}\right)_{7} = \left(\mathbf{F}^{*}\right)_{7} \; ; \; \left(\mathbf{F}_{C1}^{*}\right)_{8} = \left(\mathbf{F}^{*}\right)_{8} \; ; \quad (34)$$
$$\left(\mathbf{G}_{C1}^{*}\right)_{6} = \left(\mathbf{G}^{*}\right)_{6} \; ; \; \left(\mathbf{G}_{C1}^{*}\right)_{7} = \left(\mathbf{G}^{*}\right)_{7} \; ; \; \left(\mathbf{G}_{C1}^{*}\right)_{8} = \left(\mathbf{G}^{*}\right)_{8} \; ;$$

Here $\mathbf{U}^{*}$, $\mathbf{F}^{*}$ and $\mathbf{G}^{*}$ are obtained from the two-dimensional HLL Riemann solver. These are the only differences that arise when formulating an HLLC Riemann solver for MHD. The rest of the formulae in the previous section directly carry over to MHD. The *z*-component of the electric field is obtained by following the plan in Balsara & Spicer [10] and is given by

$$E_z = \frac{1}{2}\left(\left(\mathbf{F}^*_{C1}\right)_7 - \left(\mathbf{G}^*_{C1}\right)_6\right) \tag{35}$$

This, therefore, completes our description of the two-dimensional HLLC Riemann solver for MHD.

## IV) A Predictor-Corrector Type Numerical Scheme that Implements the Two-dimensional HLLC Riemann Solver

### IV.a) Some Caveats on the Construction of Facial Fluxes from the Edge-Centered Fluxes

The HLLC Riemann solver presented here is inherently two-dimensional and returns two fluxes whenever it is applied to one edge between four zones. Thus it might seem natural that each face in a three-dimensional calculation should take a one-fourth contribution to its flux from the four edges that bound it. However, because the conservation laws are being solved on an underlying computational mesh, the interplay between the mesh and the Riemann solver has to be considered. It is, therefore, worth making a few useful points about the two-dimensional HLLC Riemann solver that has been presented here.

1) Say we have a very slow moving contact discontinuity in a sub-sonic flow. As a result, $S^*_{Mx}$ and $S^*_{My}$ are very small (or zero) and we also have $S_L \leq 0 \leq S_R$ and $S_D \leq 0 \leq S_U$. In such situations, the two-dimensional HLLC Riemann solver used by itself can produce extremely sharp, almost stationary profiles on the computational mesh. As a result, it is always best to blend in a small fraction of an HLL solution into the HLLC solution. Extensive experimentation has shown that blending ~85% of the flux from eqns. (20) and (25) along with a complementary percentage of the flux from eqns. (10) and (13) always stabilizes such profiles.

2) When the flow is supersonic in both directions, the multidimensional Riemann solver can contribute a flux to a face that is not obtained from either of the zones that abut that face. To consider an example, if $S_L \geq 0$ and $S_D \geq 0$ then we have $\mathbf{F}^* = \mathbf{F}_{LD}$ and $\mathbf{G}^* = \mathbf{G}_{LD}$. In that case, the lower edges of the *x*- and *y*-faces will contribute fluxes to those faces that are drawn exclusively from zones that do not come together at those faces. Taken to its extreme, this would result in an even-odd decoupling instability. This is a trait that plagues all fully multidimensional Riemann solvers. While elaborate solutions may be devised, a simple fix is adequate here. It consists of realizing that we do have the results from the one-dimensional Riemann problems at the edge. Thus we blend in a fraction of the flux from the one-dimensional Riemann problem along with a complementary fraction from the multi-dimensional Riemann solver. The one-dimensional flux can always be chosen so as to restore coupling between zones that come together at the face in question. (I.e. the one dimensional Riemann problem that we pick is the one that corresponds to the face in question. This does not increase the cost of the scheme because the one-dimensional Riemann solvers are always built at each edge in the process of building the multidimensional Riemann solver.) While this degrades the maximal CFL number, it is a simple fix-up that always works. In practice, we have found that blending 40% to 60% of the flux from the one-dimensional Riemann solver with a complementary amount of flux from the multidimensional Riemann solver always restores stability while retaining a robust CFL number.

3) For evaluating the electric field in MHD, one indeed desires a fully multidimensional contribution. For that reason, when using this Riemann solver for MHD, the blending mentioned in the above point is not used for the electric fields.

4) Another way of overcoming the even-odd decoupling that was mentioned in point 2) above would be to use a small amount of Lapidus viscosity. Indeed, a Lapidus viscosity of 0.1 to 0.2 was found to work in all cases. Because the addition of such numerical viscosities can also produce unwanted dissipation, it was not used in any of the examples presented in this paper. This point does, however, hold out the possibility that a CFL number of unity can be achieved in two dimensions if one is willing to include a modest amount of Lapidus viscosity.

5) In a few rare instances one may have a triple point that is rather stationary relative to the mesh. At the location of the triple point, there is clearly no single, unique direction for the density gradient that can be provided to the multidimensional HLLC Riemann solver; besides all choices of the density gradient are poor choices. The present multidimensional HLLC Riemann solver, without any modifications, will stably integrate such situations. However, one may want to avoid asserting a direction for the density gradient when all such choices are of a single direction are ill. The gradient of the density at the edge being considered can then be compared with the similar gradient at its nearest neighbor edges. If the density gradients are reasonably aligned, there is no need to do anything and the unmodified, multidimensional HLLC Riemann solver is used. If they are strongly misaligned, one may revert to the multidimensional HLL Riemann solver. This point is not a critical issue in the design of the multidimensional HLLC Riemann solver. In the examples shown below, any agreement within $20^o$ in neighboring gradients was taken to imply perfect alignment; an angle in excess of $50^o$ between neighboring gradients was taken to imply full misalignment.

**IV.b) Construction of a Second Order Accurate Predictor-Corrector Scheme Using the Multidimensional Riemann Solver**

To derive the maximal benefit from our two-dimensional HLLC Riemann solver, we briefly catalogue a second order accurate predictor-corrector type scheme that relies exclusively on the two-dimensional HLLC Riemann solver. I.e., the predictor step uses the two-dimensional HLLC Riemann solver, and it is also used in the corrector step. Because of its inherent isotropy, the two-dimensional HLLC Riemann solver restores a much-needed isotropy to the overall simulation when used in the corrector step. Notice that when used in this fashion, the present Riemann solver is called twice at each edge during every time step. This increases the cost of the scheme but, in turn, it yields a scheme with a larger CFL number.

Since the CFL number of the scheme is controlled by the predictor step, our use of the multidimensional Riemann solver in the predictor step has the happy consequence of raising the maximal CFL number. The scheme is general enough to be used for three-dimensional flow simulations, where we have empirically found a maximal CFL number of about 0.6. A CFL number of ~0.7 was routinely used for two-dimensional problems, with some two-dimensional problems running at larger CFL numbers.

The previous two paragraphs also suggest another useful variant of the present scheme. It consists of keeping the same predictor step as in a conventional second order Godunov scheme. (I.e., the piecewise linear slopes in the flow variables within each zone are used at the start of the time step to construct the time evolution of those variables within that zone.) There is no improvement in the CFL number when this choice is made. The two-dimensional HLLC Riemann solver can then be used exclusively in the corrector step, with the goal of introducing more isotropy into the flow solver. It is indeed possible to design such a scheme and it has better handling of multidimensional flow features. Notice that when used in this fashion, the present Riemann solver is called only once at each edge during every time step. Such schemes do about the same number of float point operations as a conventional second order Godunov scheme and cost about the same too.

The way the multi-dimensional Riemann solver is implemented in the predictor and corrector steps is about the same. Spatial differencing of the predicted fluxes provides an adequate strategy for obtaining the predicted time rate of change of the flow variables. In keeping with our space-time view, the predictor step can be viewed as a step that reconstructs the piecewise linear time variation within a zone, just as the limiters reconstruct the piecewise linear spatial variation within a zone. The only difference between predictor and corrector steps is that the corrector step also uses the time rate of change in the variables, and that is provided by the predictor step. Thus the corrector step builds space- and time-centered fluxes while the predictor step only builds fluxes that are centered in space but defined at the beginning of the time step.

In this paragraph we describe how the multidimensional Riemann solver is implemented on a three-dimensional mesh, since a two-dimensional implementation is indeed trivial. The processing is done in sweeps that are aligned with the edges of the mesh, and the individual sweeps are described below.

Z-Sweep: When sweeping through the *z*-edges of a three-dimensional mesh, one simply uses the reconstructed variables to obtain the four states that come together at that edge. The Riemann solver is then used as a machine that accepts four states as its inputs and produces two fluxes in the *x*- and *y*-directions as its outputs at that particular edge. The two fluxes from that edge are then used to make a one-fourth contribution to the flux variables at the four faces that come together at that edge. Thus when sweeping through the *z*-edges, the *x*- and *y*-fluxes take on contributions. The caveats from Sub-section IV.a have to be respected. If the problem is an MHD problem, the present Riemann solver will have directly yielded the *z*-component of the edge-centered electric field via eqn. (35). The *z*-sweep is described first because the narrative in this paper is built around processing *z*-edges. Only one version of the present Riemann solver needs to be written. The *x*- and *y*-sweeps will be described next in such a way that they reuse this Riemann solver.

Y-Sweep: Now we sweep through the *y*-edges. The four states that come together at each *y*-edge are again obtained from the reconstructed variables. The only change is to make a passive rotation of the axes so that the *y*-velocity is mapped to the *z*-velocity; the *z*-velocity is mapped to the *x*-velocity and the *x*-velocity is mapped to the *y*-velocity. This simple permutation is applied to all the vector variables in the flow. The states are then fed into the Riemann solver which now returns fluxes in the *x*- and *z*-directions. The vector components of those fluxes have to be permuted back, so as to undo the prior permutation. Again, one-fourth of the *x*- and *z*-fluxes at the *y*-edge are added into the facial fluxes. The suggestions from Sub-section IV.a have to be used. In an MHD problem, the *y*-component of the edge-centered electric field is now fully built.

X-Sweep: Now we sweep through the *x*-edges. The four states that come together at each *x*-edge are again obtained from the reconstructed variables. The only change is to make a passive rotation of the axes so that the *x*-velocity is mapped to the *z*-velocity; the *y*-velocity is mapped to the *x*-velocity and the *z*-velocity is mapped to the *y*-velocity. This simple permutation is applied to all the vector variables in the flow. The states are then fed into the Riemann solver which now returns fluxes in the *y*- and *z*-directions. The vector components of those fluxes have to be permuted back, so as to undo the prior permutation. Again, one-fourth of the *y*- and *z*-fluxes at the *x*-edge are added into the facial fluxes. The ideas from Sub-section IV.a have to be utilized. By this point, the flux in each face will have taken a one-fourth contribution from each of the four edges that surround it. In an MHD problem, the *x*-component of the edge-centered electric field is now fully built.

This completes our description of how the two-dimensional HLLC Riemann solver is implemented in code.

It is important to compare the speed of the present code with the speed of conventional, second order Godunov schemes. When we used a conventional, TVD scheme for Euler flow we were able to update 212,600 three-dimensional zones per second. This scheme relied on a one-dimensional linearized Riemann solver and had a maximal CFL number of about 0.3. When the multidimensional Riemann solver was used in the predictor and corrector steps, we updated 90,400 three-dimensional zones per second and sustained a maximal CFL number of ~0.6. When the multidimensional Riemann solver was used exclusively in the corrector step, we were able to update 166,300 three-dimensional zones per second and sustained a maximal CFL number of ~0.3. This timing study used a single core of a modern Intel Xeon processor operating at 3 GHz. The multidimensional Riemann solver-based technology, therefore, shows itself to be cost-competitive with the conventional one-dimensional Riemann solver technology while delivering a superior result.

For three-dimensional numerical MHD, where the reconstruction as well as the Riemann solver is more costly, a conventional scheme with a one-dimensional linearized Riemann solver updates 121,500 zones per second and operates with a maximal CFL number of about 0.3 . The scheme that uses the multidimensional Riemann solver in both predictor and corrector steps updates 56,700 zones per second and operates with a maximal CFL number of ~0.6. A scheme that uses the multidimensional Riemann solver exclusively in the corrector step (but not in the predictor step) updates 95,000 zones per second and operates with a maximal CFL number of ~0.3. These numbers were obtained on a single core of a modern Intel Xeon processor operating at 3 GHz. We see, therefore, that the two-dimensional HLLC Riemann solver technology is cost-competitive with the conventional second order Godunov scheme technology for MHD simulations. Besides, as the next section will show, it delivers superior solutions.

## V) Results

In Balsara [7] we presented an accuracy analysis of second order Godunov schemes that use the multidimensional Riemann solver technology. In that work we showed that the algorithm designed in the previous section meets second order accuracy. For that reason, we do not repeat it here. In this Section we present several test problems which highlight the advantages of a multidimensional Riemann solver approach. All the tests are, by default, based on using the multidimensional Riemann solver in both the predictor and corrector steps. As a result, the CFL numbers will be larger than those that are permitted by conventional second order Godunov schemes that use one-dimensional Riemann solver technology. In Sub-section V.a we present several test problems involving the Euler equations. Sub-section V.b presents test problems for the MHD equations.

## V.a) Hydrodynamic Tests

Several interesting hydrodynamical test problems are presented, each of which shows some of the advantages of introducing multidimensional Riemann solvers in current flow solver technology.

### V.a.1) Spherical Blast Wave in Three Dimensions

It is well-known that higher order Godunov codes (including several that achieve better than second order accuracy) produce anisotropic explosions. This trend is usually exacerbated when the simulation is starved of numerical resolution. Ironically, this is the usual way in which these codes are used in scientific and engineering applications. For that reason, we picked a small $64^3$ zone mesh that spans the three-dimensional domain $[-0.5, 0.5]^3$. The boundaries were continuative. A stationary fluid with a unit density and a polytropic index of 1.4 was initialized on the mesh. The pressure was set to 0.1 all over, except in a sphere of radius 0.1, where it was set to 1000. The simulation was run to a final time of 0.02.

Figs. 6a shows the density in the midplane when a conventional second order Godunov scheme was used and Fig. 6b shows the same information when the scheme from the previous section was used. An MC limiter was used in both simulations. A conventional second order TVD scheme that was based on a one-dimensional HLLC Riemann solver was run with a CFL number of 0.3 to obtain the results in Fig. 6a. The two-dimensional Riemann solver-based scheme with a CFL of 0.6 was used for the results in Fig. 6b. We clearly see that the density in Fig. 6b is more isotropic than the density in Fig. 6a. To make the point more clearly, we plot out the density along the x-axis (solid line) and along a $45^o$ direction (dashed line). Fig. 6c shows such a plot for the data in Fig. 6a. Fig. 6d shows such a plot for the data in Fig. 6b. We clearly see that the density from the scheme that used the multidimensional Riemann solver is much more isotropic than the density from the conventional second order Godunov scheme.

It is also possible to use the multidimensional Riemann solver exclusively in the corrector step. In that case, the predictor step is the conventional one that is used in most second order Godunov schemes. The resulting scheme will have a useful CFL limit of ~0.3 in three dimensions. We have, nevertheless, verified that the resulting solution to this test problem is as good as the results shown in Figs. 6b and 6d. This shows that the two-dimensional HLLC Riemann solver can also be used to design a low cost scheme with superior propagation properties for flow features of interest.

**V.a.2) Multidimensional Riemann Problems in Two Dimensions**

Schulz-Rinne, Collins and Glaz [47] showed the value of using multidimensional Riemann problems for calibrating numerical schemes. Brio, Zakharian and Webb [14] provided explicit values for the initial conditions for some of these Riemann problems, and we catalogue them here. The first multidimensional Riemann problem consists of setting

$\rho = 0.5313$,    P=0.4,    $v_x = 0.0$,    $v_y = 0.0$    for x>0, y>0

$\rho = 1.0$,    P=1.0,    $v_x = 0.0$,    $v_y = 0.7276$    for x>0, y<0

$\rho = 1.0$,    P=1.0,    $v_x = 0.7276$,    $v_y = 0.0$    for x<0, y>0

$\rho = 0.8$,    P=1.0,    $v_x = 0.0$,    $v_y = 0.0$    for x<0, y<0.

The problem initially starts off as two weak shocks and two slip lines. The problem was run with a CFL number of 0.9 on a 1000×1000 zone mesh that spans $[-1,1]^2$. It was stopped at a time of 0.601. The piecewise linear part of an r=3 WENO reconstruction was used along with the multidimensional Riemann solver designed here. The density variable at the latest time in this problem is shown in Fig. 7a. The final solution, which is dominated by the region of strong interaction, can be interpreted as two Mach reflections and two contact surfaces at the intersection of the four shocks. We see that a very pronounced density valley moves to the intersection point of the four shocks, which is in keeping with expectations. It has been suspected that the Mach stem can become Kelvin-Helmholtz unstable. Indeed, by running this problem on a larger mesh with a conventional scheme, one can see this instability emerging. The very interesting point being made here, however, is that we can see this instability emerge on a much smaller mesh when the effects of true multidimensionality, coupled with an ability to track sub-structure, are incorporated into the Riemann solver.

The next multidimensional Riemann problem consists of setting

$\rho = 1.5$,    P=1.5,    $v_x = 0.0$,    $v_y = 0.0$    for x>0, y>0

$\rho = 0.5323$,    P=0.3,    $v_x = 0.0$,    $v_y = 1.206$    for x>0, y<0

$\rho = 0.5323$,    P=0.3,    $v_x = 1.206$,    $v_y = 0.0$    for x<0, y>0

$\rho = 0.1379$,    P=0.029,    $v_x = 1.206$,    $v_y = 1.206$    for x<0, y<0.

The problem results in a double Mach reflection and a shock propagating at $45^o$ to the mesh. The problem was run with a CFL number of 0.9 on a 1000×1000 zone mesh that spans $[-1,1]^2$. It was stopped at a time of 1.1. The piecewise linear part of an r=3 WENO reconstruction was used along with the multidimensional Riemann solver designed here. The density variable from the lower left quadrant of the simulation is shown in Fig. 7b; this is also the region where all the waves interact strongly with one another. We see that the mushroom cap is captured very crisply in this problem owing to the use of the multidimensional Riemann solver. At the resolution used for this problem, one can begin to trace out the emergence of a Kelvin-Helmholtz instability. As in Fig. 7a, the inclusion of multidimensional effects and sub-structure in the Riemann solver, show their obvious uses.

### V.a.3) Double Mach Reflection Problem

Woodward and Colella [51] proposed this test problem, and we use the same parameters and set-up as those authors. The problem was run to a time of 0.2 using a CFL number of 0.7. As is customary, we only image the domain $[0, 3]\times[0, 1]$. Figs. 8a and 8b show the density variables at the final time at $1920\times480$ and $2400\times600$ zone resolution. The two smaller panels at the bottom of the figure, i.e. Figs. 8c and 8d, show a blow-up of the region around the Mach stem from the two former computations. The piecewise linear part of an r=3 WENO reconstruction was used along with the multidimensional Riemann solver. Notice that our scheme resolves all the structures, including the instabilities that develop around the Mach stem. Cockburn & Shu [16] showed that they needed a fourth order scheme operating at $1920\times480$ zone resolution to capture the Kelvin-Helmholtz instability of the Mach stem. Despite being second order, the $1920\times480$ zone simulation in Fig. 8 has already picked up the Kelvin-Helmholtz instability at the Mach stem, and the $2400\times600$ zone simulation only improves on it. We see, therefore, that the inclusion of true multidimensionality has enabled the second order schemes to almost catch up with their higher order cousins. This also makes a compelling case for including multidimensional effects in Godunov schemes that go beyond second order (Jiang & Shu [33], Balsara & Shu [9]), a task that will be undertaken later.

### V.b) MHD Tests

We present several stringent MHD test problems in this sub-section. They were all carried out using a divergence-free MHD scheme that obtained the multidimensionally upwinded electric field by using the suggestion of Balsara & Spicer [10] along with the multidimensional Riemann solver methods presented here. Please note that none of the tests involving the multidimensional Riemann solver required a doubling of the dissipation (Londrillo and DelZanna [36], Gardiner & Stone [28]) in order to obtain the edge-centered electric fields.

### V.b.1) Long Term Decay of Alfven Waves in Two Dimensions

This problem studies the numerical dissipation of the scheme by examining the long term amplitude decay of torsional Alfven waves. It was originally presented in Balsara [5]. We do not describe it here, except to mention that in this problem an Alfven wave is propagated around a completely periodic, two-dimensional domain in the *xy*-plane for a very long period of time. A good MHD algorithm should propagate the Alfven wave with minimal decay, because that is the analytical solution of the MHD equations. In practice, the amplitudes of the *z*-velocity and the *z*-magnetic field do indeed decay because of the numerical dissipation that is inherent in any numerical scheme. The extent of the decay is a qualitative reflection of the numerical dissipation in the code. (Because all the schemes used here are nonlinearly stabilized, it is not possible to use this problem to quantify the numerical dissipation.)

Fig. 9 presents log-linear graphs that show the decay in the amplitude of the Alfven wave with time. Figs. 9a and 9b show the *z*-velocity and *z*-magnetic field respectively when different schemes are used. The solid and dotted curves show the decay in the Alfven wave when the multidimensional Riemann solver is used along with the piecewise linear part of an r=3 WENO limiter and the MC limiter respectively. The dashed and dot-dash curves show the decay in the Alfven wave when a one-dimensional HLLC Riemann solver is used in a conventional Godunov scheme with the same two reconstruction strategies quoted previously. Because the WENO reconstruction is very smooth, we see that it produces a very smooth decay in the Alfven wave. The smoothness also gives the WENO results a smaller decay. The MC limiter shows a small amount of jitter owing to the fact that the limiter abruptly switches directions. We see too that for a specified reconstruction strategy, the multidimensional Riemann solver produces less dissipative simulations than the one-dimensional Riemann solver. The simulations that use the multidimensional Riemann solvers were run with a CFL number of 0.7; the other simulations used a CFL number of 0.45.

**V.b.2) Field Loop Advection in Two Dimensions**

This problem is set up on a 128x64 zone domain that spans [-1,1]×[-0.5,0.5]. The problem consists of advecting a two-dimensional loop of magnetic field with a very low magnetic pressure compared to the gas pressure. The loop is advected along the diagonal of the computational domain with periodic boundaries applied in both directions. It was initially described in Gardiner & Stone [28] and we do not repeat the description here. The conventional result examines the structure of the magnetic loop after it has completed one complete orbit around the domain. We present the results of running this calculation twice, with competing algorithms. First, the problem was run with an MC limiter and the multidimensional Riemann solver technology described in this paper. A CFL number of 0.9 was used and no doubling of the numerical dissipation was seen to be necessary when evaluating the edge-centered electric fields. By way of comparison, we subsequently ran the same problem with an MC limiter and a linearized, one-dimensional Riemann solver so that all the waves in each direction could be captured. The prescription for doubling the dissipation in the electric fields, as described in Gardiner & Stone [28], was implemented in the linearized Riemann solver. This latter calculation was run with conventional higher order Godunov scheme technology with a CFL number of 0.45.

Figs. 10a and 10b show the magnitudes of the magnetic field for the field loop advection problem after it has executed one complete orbit around the computational domain. Fig. 10a is based on conventional second order Godunov methods with the Gardiner & Stone [28] fix for the electric field. Fig. 10b presents the results from the multidimensional HLLC Riemann solver-based scheme. Fig. 10b shows that the loop's profile is extremely isotropic, owing to the use of the multidimensional Riemann solver technology. Despite the lower CFL number and the use of a seemingly better Riemann solver (i.e. a linearized Riemann solver v/s an HLLC Riemann solver), we see that the results in Fig. 10a are clearly inferior to the results in Fig. 10b. We see from Fig. 10b that the numerical diffusion of the loop's boundaries is minimal and there are no oscillations of the magnetic pressure within the loop. This result also makes a strong case for introducing multidimensional Riemann solver technology in Godunov schemes for MHD that go beyond second order (Balsara [4], Balsara *et al* [8], Balsara [5]). We will undertake such a task later.

It is also possible to use the multidimensional Riemann solver developed here exclusively in the corrector step. In that case, the predictor step is the conventional one that is used in most second order Godunov schemes for MHD. The resulting scheme will have a useful CFL limit of ~0.45 in two dimensions. We have, nevertheless, verified that the resulting solution to this test problem is as good as the results shown in Fig. 10b. This shows that the two-dimensional HLLC Riemann solver can also be used to design a low cost scheme for MHD with superior propagation properties for flow and magnetic field features of interest.

**V.b.3) Three Dimensional MHD Blast Wave with very low Plasma Beta**

This problem is a three-dimensional MHD extension of the problem proposed in Balsara & Spicer [10]. The computational domain remains the unit cube, and the MHD blast problem is solved on a computational mesh of $96^3$ zones spanning $[-0.5, 0.5]^3$. The initially stationary gas has unit density and $\gamma = 1.4$. A pressure pulse is initialized within a spherical region with radius 0.1. The interior and exterior gas pressures are 1000 and 0.1, respectively. Unlike the hydrodynamic blast problem, we include a magnetic field threading the entire domain which initially points along the diagonal of the box, given by $(B_x, B_y, B_z) = (70/\sqrt{3}, 70/\sqrt{3}, 70/\sqrt{3})$. Note that this magnetic field corresponds to a plasma-$\beta$ of 0.000513 in the ambient medium. The problem was run to a time of 0.014 with a CFL number of 0.6 using the multidimensional Riemann solver-based scheme catalogued here.

Figs. 11a through 11d show the density, pressure, magnitude of the velocity and magnitude of the magnetic field respectively in the midplane of the simulation. Despite the very low plasma-beta, the pressure remains robustly positive. As the blast propagates, large velocities develop and the magnetic field is compressed so as to become even stronger than its initial value. The pressure is obtained by subtracting the kinetic and magnetic energies from the total energy density. Conventional second order Godunov schemes produce a negative pressure in MHD simulations where the velocity and magnetic field become large. Fig. 11b shows that this is not so for the MHD scheme presented here. We see, therefore, that along with the large time steps, the multidimensional Riemann solver-based methods also do much better at maintaining pressure positivity. This stems from their ability to propagate magnetic fields very accurately in any needed direction on the computational mesh.

**VI) Conclusions**

A genuinely two-dimensional HLLC Riemann solver has been presented. The Riemann solver is implemented at edges, where it accepts the four states from the four surrounding zones as inputs. The output of the HLLC Riemann solver consists of two fluxes, one in each direction. In this work we have addressed the issues that arise when one wants to incorporate a contact discontinuity into such a multidimensional Riemann solver. Just as the one-dimensional HLLC Riemann solver is obtained by introducing sub-structure into the one-dimensional HLL Riemann solver, the multidimensional HLLC Riemann solver is also obtained from the multidimensional HLL Riemann solver. The orientation of the contact discontinuity has to be provided as an input, though it can propagate in any direction and with any strength based on the local structure of the flow. A principal advance in this work is that we show that all the space-time integrals that need to be done in defining such a two-dimensional Riemann solver can be reduced to forms that are seamlessly, and easily, implemented on a computer. The method is easily extensible to any system of conservation laws.

The process of obtaining face-centered fluxes is described in some detail. A second order method for conservation laws that incorporates the present Riemann solver in the predictor and corrector steps is also described. It is shown to permit larger CFL numbers for two- and three-dimensional problems than those used for conventional second order Godunov schemes. If the multidimensional Riemann solver is used only in the corrector step, the CFL number is the same as that for any traditional higher order Godunov scheme and its cost per time step is also competitive.

The scheme is shown to work very well for several difficult hydrodynamical problems. Isotropic propagation of flow features is one of the strengths of the multidimensional Riemann solver technology developed here. The ability to propagate contact discontinuities with fidelity is another one of its strengths.

The versatility of the present multidimensional HLLC Riemann solver is also demonstrated by extending it to the MHD system. Because the present Riemann solver is applied at edges, it naturally yields the multidimensionally upwinded electric fields at zone-edges. These electric fields then enable one to make a divergence-free update of the magnetic field without resort to any doubling of the dissipation. The method is shown to work well for several stringent MHD test problems, including very low plasma-beta problems where the positivity of the pressure variable would otherwise become an issue. Owing to its inherent multidimensionality, the method also shows superior propagation of magnetic field structures in any direction on the computational mesh.

## Acknowledgements

The author acknowledges support via NSF grant NSF-AST-0947765 and NASA grant NASA-NNX08AG69G. The majority of simulations were performed on a cluster at UND that is run by the Center for Research Computing. The author thanks C. Meyer and C. Matthews for help with plotting the results.

**Appendix:**

We can also illustrate the integration over the lower triangular face for the situation where $S_L \leq X_D \leq S_R$ and $S_L \leq X_U \leq S_R$ and $X_D S_U - S_D X_U < 0$. For the lower triangular face, the areal integration extends over the triangle with space-time vertices given by $(0,0,0)$, $(S_R T, S_D T, T)$ and $(X_D T, S_D T, T)$; where the right hand rule would produce an outward pointing normal if the three points were traversed in sequence. We refer to this as a *Type II integration over the lower triangular face*, and we point out that it corresponds to the range $x \in [X_D, S_R]$ of the lower boundary in Fig. 5. The task of evaluating the area integral reduces to a matter of figuring out where $X_D$ is located relative to the wave speeds at the lower face. We present pseudocode for carrying out the area integral over the lower triangular face of the inverted pyramid in Fig. 3b below. An obvious factor of $T^2$ has been removed from the area integral. The limits of integration over each sub-triangle that corresponds to a constant state are denoted $X_1$ and $X_2$ below, where we have $X_2 > X_1$.

$(\text{Sums over triangles in the lower face}) = 0$

$if \left( X_D < S_L^D \right) \{$

$\quad X_1 = \max(X_D, S_L) \; ; \; X_2 = S_L^D \; ;$

$\quad (\text{Sums over triangles in the lower face}) + = \dfrac{(X_2 - X_1)}{2} \left( -\mathbf{G}_{LD} + S_D \mathbf{U}_{LD} \right) \; ;$

$\}$

$if \left( X_D < S_M^D \right) \{$

$\quad X_1 = \max(X_D, S_L^D) \; ; \; X_2 = S_M^D \; ;$

$\quad (\text{Sums over triangles in the lower face}) + = \dfrac{(X_2 - X_1)}{2} \left( -\mathbf{G}_D^{*-} + S_D \mathbf{U}_D^{*-} \right) \; ;$

$\}$

$if \left( X_D < S_R^D \right) \{$

$\quad X_1 = \max(X_D, S_M^D) \; ; \; X_2 = S_R^D \; ;$

$\quad (\text{Sums over triangles in the lower face}) + = \dfrac{(X_2 - X_1)}{2} \left( -\mathbf{G}_D^{*+} + S_D \mathbf{U}_D^{*+} \right) \; ;$

$\}$

$if \left( X_D < S_R \right) \{$

$\quad X_1 = \max(X_D, S_R^D) \; ; \; X_2 = S_R \; ;$

$\quad (\text{Sums over triangles in the lower face}) + = \dfrac{(X_2 - X_1)}{2} \left( -\mathbf{G}_{RD} + S_D \mathbf{U}_{RD} \right) \; ;$

$\}$

This completes our description of Type II integration at the lower triangular face.

When we have $S_D \leq Y_L \leq S_U$ and $S_D \leq Y_R \leq S_U$ and $S_R Y_L - Y_R S_L \leq 0$, we do not integrate over the lower triangular face at all. We refer to this as *Type 0 integration over the lower triangular face*. When we have $S_D \leq Y_L \leq S_U$ and $S_D \leq Y_R \leq S_U$ and $S_R Y_L - Y_R S_L > 0$, we integrate over the entire lower triangular face. For the lower triangular face, the areal integration extends over the triangle with space-time vertices given by $(0,0,0)$, $(S_R T, S_D T, T)$ and $(S_L T, S_D T, T)$; where the right hand rule would produce an outward pointing normal if the three points were traversed in sequence. We refer to this as a *Type III integration over the lower triangular face*, and we point out that it corresponds to the range $x \in [S_L, S_R]$ of the lower boundary in Fig. 5. Here, we integrate over the entire lower triangular face. We present pseudocode for carrying out the area integral over the lower triangular face of the inverted pyramid in Fig. 3b below. An obvious factor of $T^2$ has been removed from the area integral. The limits of integration over each sub-triangle that corresponds to a constant state are denoted $X_1$ and $X_2$ below, where we have $X_2 > X_1$.

$(\text{Sums over triangles in the lower face}) = 0$

$X_1 = S_L$ ; $X_2 = S_L^D$ ;

$(\text{Sums over triangles in the lower face}) + = \dfrac{(X_2 - X_1)}{2}(-\mathbf{G}_{LD} + S_D \mathbf{U}_{LD})$ ;

$X_1 = S_L^D$ ; $X_2 = S_M^D$ ;

$(\text{Sums over triangles in the lower face}) + = \dfrac{(X_2 - X_1)}{2}(-\mathbf{G}_D^{*-} + S_D \mathbf{U}_D^{*-})$ ;

$X_1 = S_M^D$ ; $X_2 = S_R^D$ ;

$(\text{Sums over triangles in the lower face}) + = \dfrac{(X_2 - X_1)}{2}(-\mathbf{G}_D^{*+} + S_D \mathbf{U}_D^{*+})$ ;

$X_1 = S_R^D$ ; $X_2 = S_R$ ;

$(\text{Sums over triangles in the lower face}) + = \dfrac{(X_2 - X_1)}{2}(-\mathbf{G}_{RD} + S_D \mathbf{U}_{RD})$ ;

This completes our description of Types 0 and III integration at the lower triangular face.

The above discussion helps us to put together a simple decision tree to determine when one or the other type of integration is to be used at any of the four triangular faces of the inverted pyramid shown in Fig. 3b. It is given by if $\left(S_L \leq X_D \leq S_R \text{ and } S_L \leq X_U \leq S_R\right)$ then
    Left-Type III;  Right-Type 0;  Down-Type I;  Up-Type I;     when $X_D S_U - S_D X_U > 0$
    Left-Type 0;   Right-Type III; Down-Type II; Up-Type II;   otherwise
else if $\left(S_D \leq Y_L \leq S_U \text{ and } S_D \leq Y_R \leq S_U\right)$ then
    Left-Type I;   Right-Type I;  Down-Type III; Up-Type 0;    when $S_R Y_L - Y_R S_L > 0$
    Left-Type II;  Right-Type II;  Down-Type 0;  Up-Type III;  otherwise
else if $\left(S_L < X_U < S_R \text{ and } S_D < Y_R < S_U\right)$ then
    Left-Type III;  Right-Type I;  Down-Type III; Up-Type I;     when $S_R S_U - Y_R X_U > 0$
    Left-Type 0;   Right-Type II;  Down-Type 0;  Up-Type II;    otherwise
else if $\left(S_L < X_D < S_R \text{ and } S_D < Y_L < S_U\right)$ then
    Left-Type I;   Right-Type 0;  Down-Type I;  Up-Type 0;     when $X_D Y_L - S_D S_L > 0$
    Left-Type II;  Right-Type III; Down-Type II; Up-Type III;  otherwise
else if $\left(S_L < X_U < S_R \text{ and } S_D < Y_L < S_U\right)$ then
    Left-Type I;   Right-Type III; Down-Type III; Up-Type II;    when $X_U Y_L - S_U S_L > 0$
    Left-Type II;  Right-Type 0;  Down-Type 0;  Up-Type I;     otherwise
else if $\left(S_L < X_D < S_R \text{ and } S_D < Y_R < S_U\right)$ then
    Left-Type 0;   Right-Type I;  Down-Type II;  Up-Type 0;     when $S_R S_D - Y_R X_D > 0$
    Left-Type III;  Right-Type II;  Down-Type I;  Up-Type III;   otherwise
end if

The above decision tree enables us to automate the integration procedure over the triangular faces of the inverted pyramid shown in Fig. 3b. Fig. 5 again proves very useful in identifying the limits of integration over the right, left and upper faces.

**Appendix B**

    For completeness, we also denote the area of the rectangle in Fig. 4 that contains the state $\mathbf{U}_{C2}^*$ by $A_{C2}^* T^2$. We then have

$$A_{C2}^* = \left(S_R - S_L\right)\left(S_U - S_D\right) - A_{C1}^*$$

and the consistency condition between states $\mathbf{U}_{C1}^*$ and $\mathbf{U}_{C2}^*$ from the HLLC Riemann solver and $\mathbf{U}^*$ from the HLL Riemann solver then becomes

$$A_{C1}^* \mathbf{U}_{C1}^* + A_{C2}^* \mathbf{U}_{C2}^* = \left(A_{C1}^* + A_{C2}^*\right) \mathbf{U}^*$$

Now that we have a procedure for obtaining $\mathbf{U}_{C1}^*$, the above consistency condition can be used to obtain the state $\mathbf{U}_{C2}^*$, if it is needed. Our Riemann solver satisfies the consistency condition by construction.

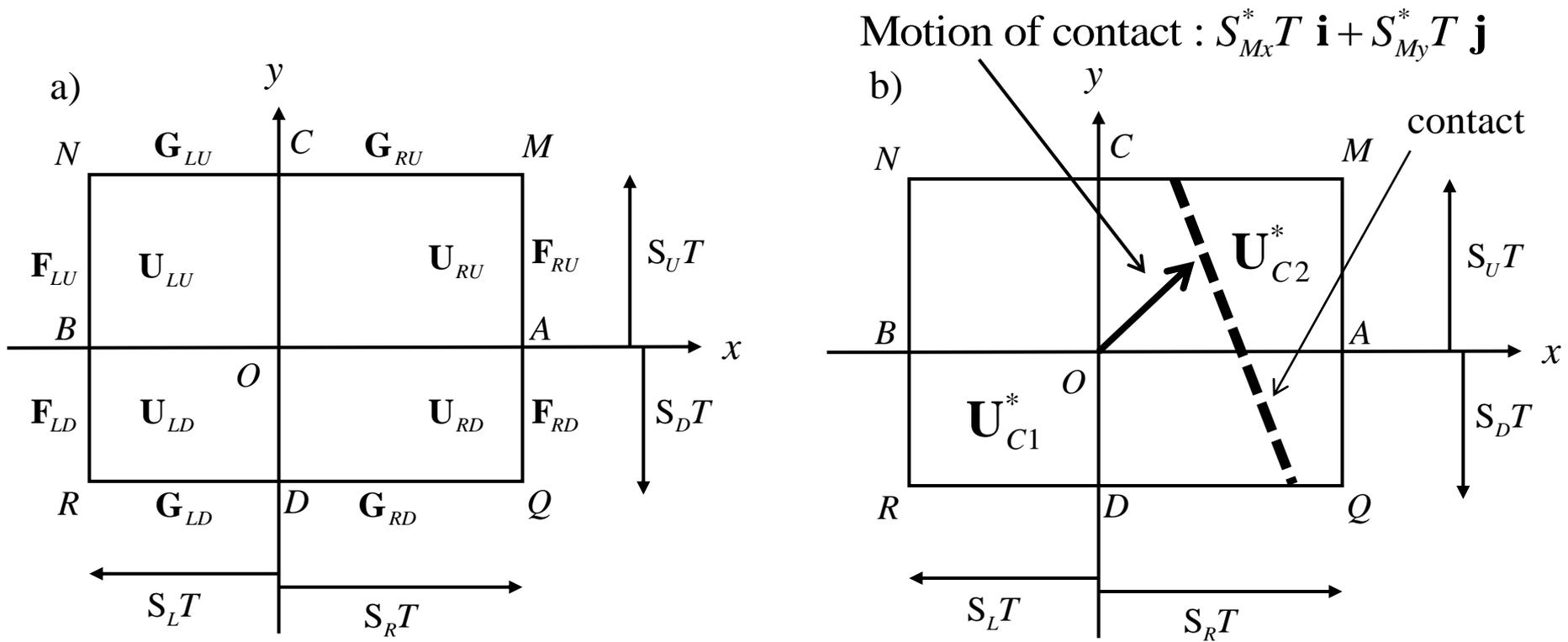

Fig. 1 depicts a situation where four neighboring zones meet at an edge. The four zones lie in each of the four quadrants of the xy-plane. The origin O of the xy-plane denotes the edge shared by the four zones. The solution vector and fluxes in the first quadrant are denoted by a subscript RU (right-up); those in the second quadrant by LU (left-up); those in the third quadrant by LD (left-down); those in the fourth quadrant by RD (right-down). The waves start propagating outward from the origin at t=0. In a time t=T, the waves propagate out to $x = S_R T$ and $x = S_L T$ along the x-axis and out to $y = S_U T$ and $y = S_D T$ along the y-axis. The rectangle QMNR bounds the domain that will be affected by the waves. Fig. 1a, on the left, shows the portion of the original domain that will be swept over by the strongly-interacting state $\mathbf{U}^*$ of the HLL Riemann solver. All the initial states and fluxes are also shown. Fig. 1b, on the right, depicts the 2D HLLC Riemann solver without repeating details of the states and fluxes. However, the contact is shown.

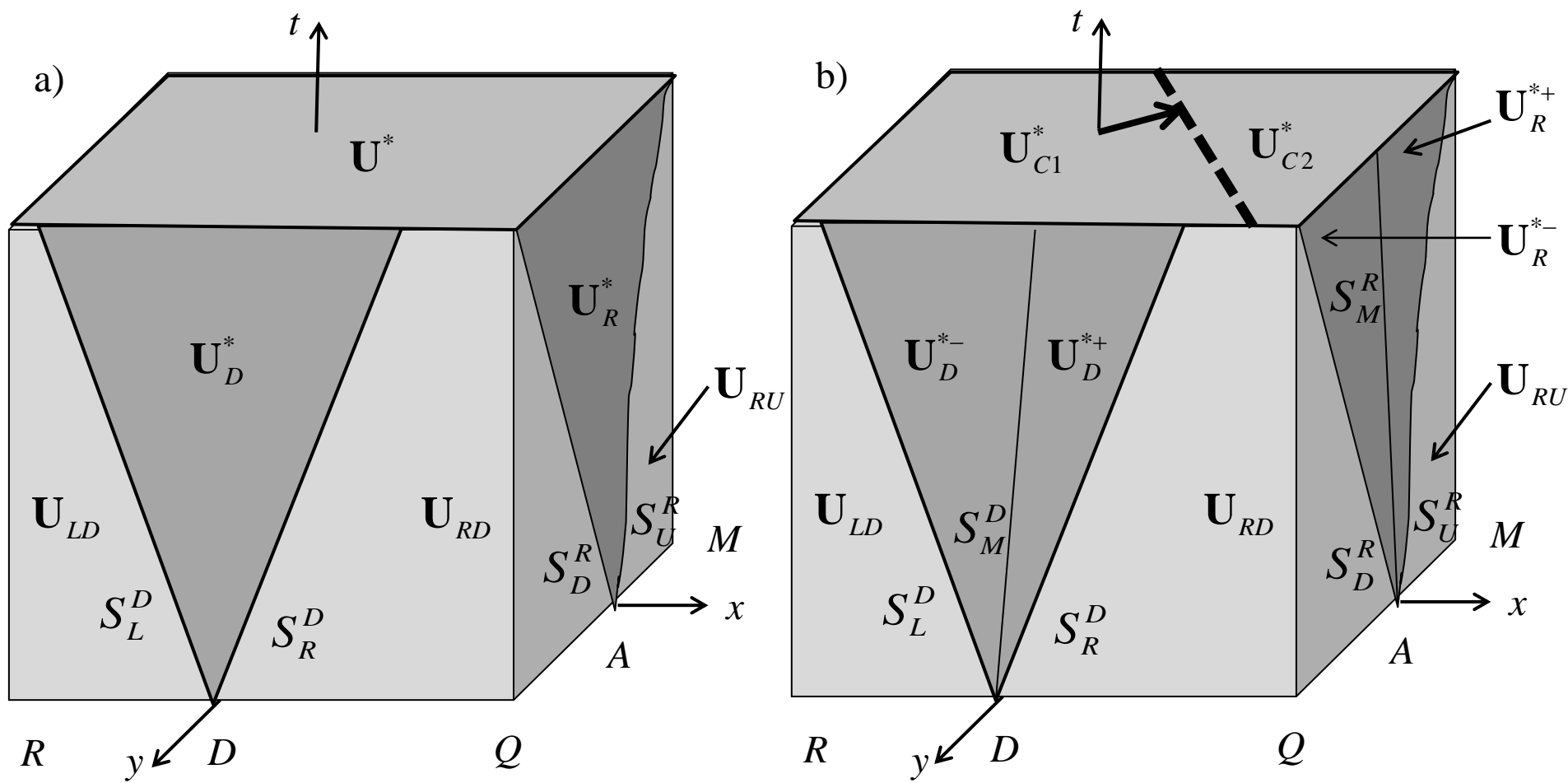

Fig. 2 shows the control volumes in space and time over which one has to integrate the two dimensional conservation law. The bases of the rectangular prisms are RQMN from Fig. 1 and the height is T. The x-t and y-t planes of the figure also show the one-dimensional Riemann problems in the right and lower faces of the control volumes. Fig. 2a, on the left, shows the two-dimensional HLL Riemann solver with only one strongly-interacting state. Fig. 2b, on the right, shows the two-dimensional HLLC Riemann solver. Notice that each of the one-dimensional Riemann problems has a contact discontinuity in the right panel. The top of the control volume in the right panel also has a multidimensional contact discontinuity separating two strongly-interacting states.

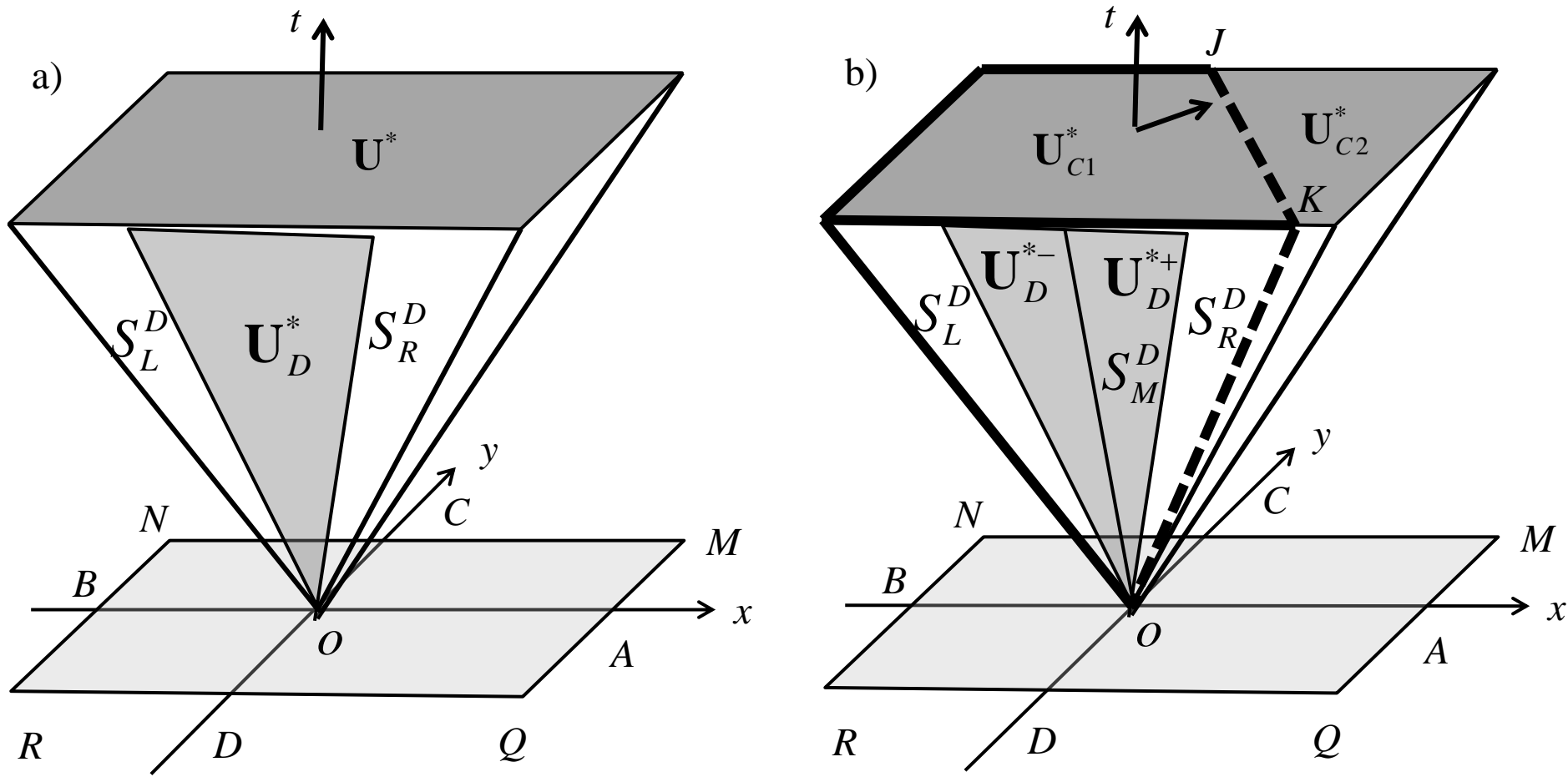

*Fig. 3 can be thought of as removing the side panels of Fig. 2 so as to expose the space-time extent of the strongly-interacting state(s). The bases of the rectangular pyramids match with RQMN from Fig. 1 and the height is T. Fig. 3a, on the left, shows the situation for the two-dimensional HLL Riemann solver. Fig. 3b, on the right, shows the situation for the two-dimensional HLLC Riemann solver. The strongly-interacting state(s) form an inverted rectangular pyramid in space-time. We also show the projection of the x-directional Riemann problem at the lower face on to the corresponding face of the pyramid. The very thick dashed line separates the strongly-interacting states $\mathbf{U}^*_{C1}$ and $\mathbf{U}^*_{C2}$.*

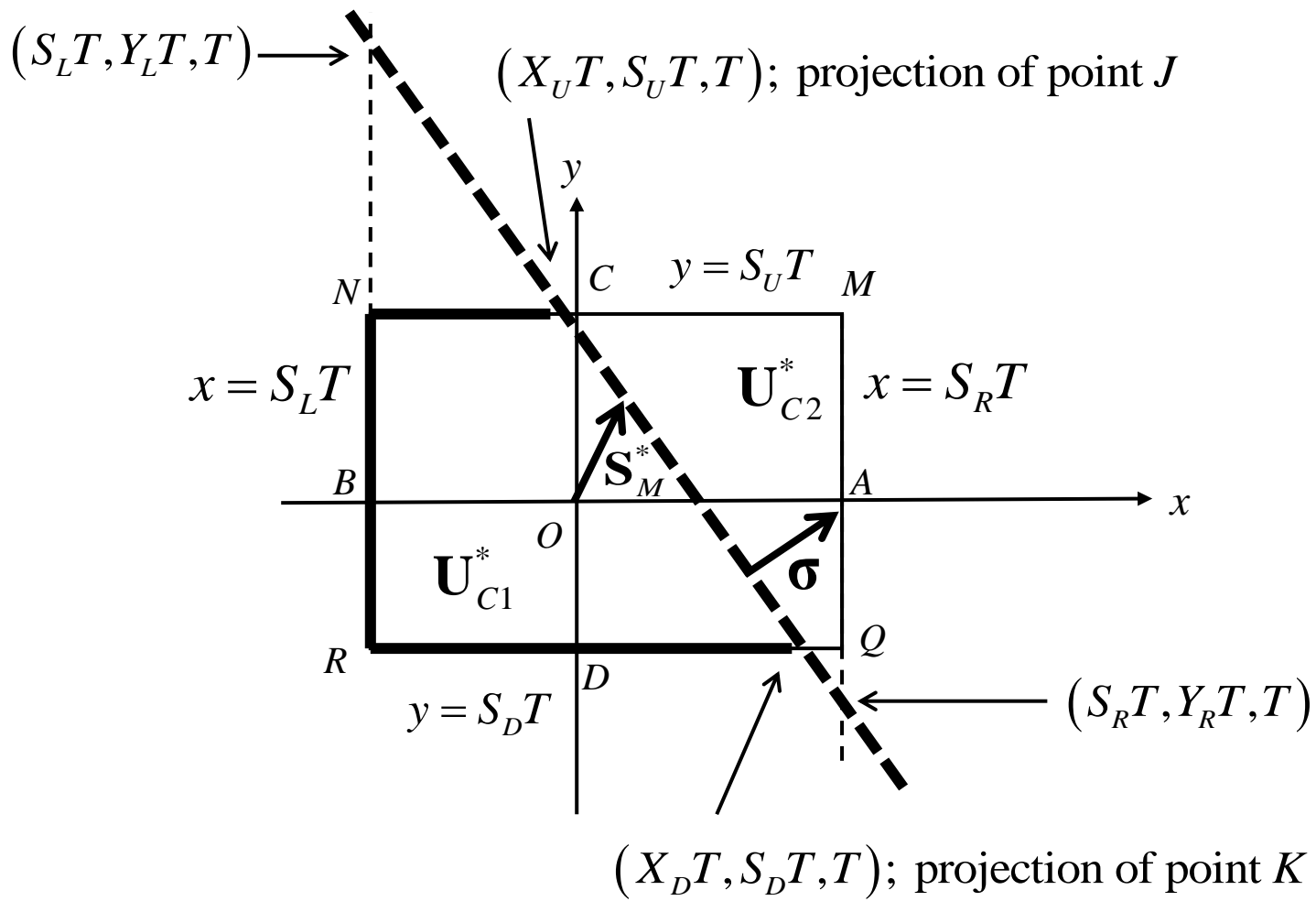

Fig. 4, which shows a projection of the rectangular base of the inverted pyramid from Fig. 3b on to the x-y plane. The very thick dashed line still shows the contact discontinuity. Along with the thick solid lines, it serves to identify the region that is of interest to us. The locations in space and time where the contact discontinuity intersects the faces of the rectangle are also shown.

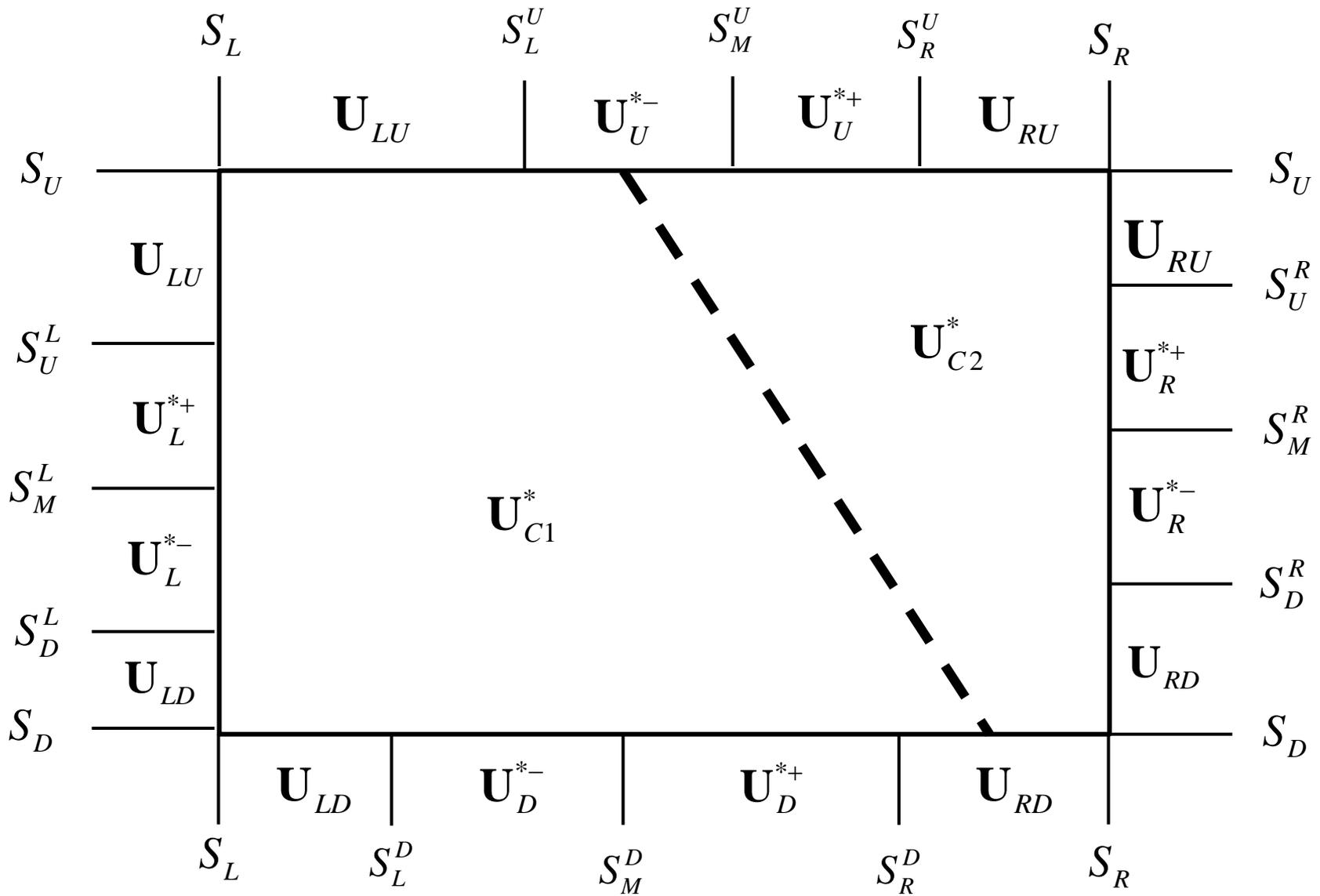

Fig. 5 is designed to be a diagram that shows us how the wave speeds are labeled and ordered on each of the faces and gives the names of the states that lie between the waves. It is very useful in identifying the limits of integration over the triangular faces of the inverted pyramid shown in Fig. 3b. It can also be used to identify the states that are to be used in the supersonic limits.

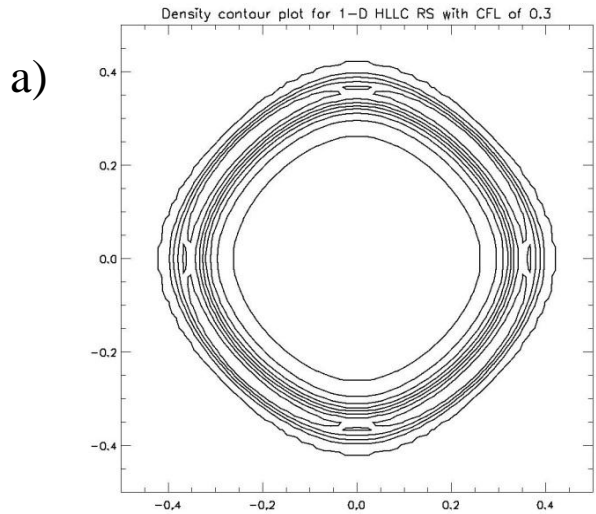 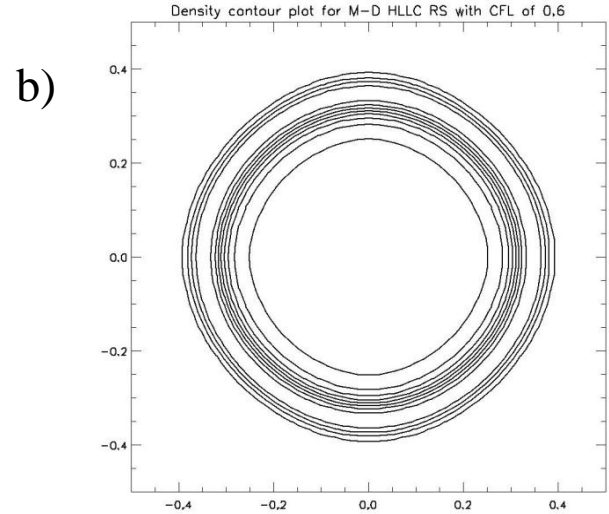
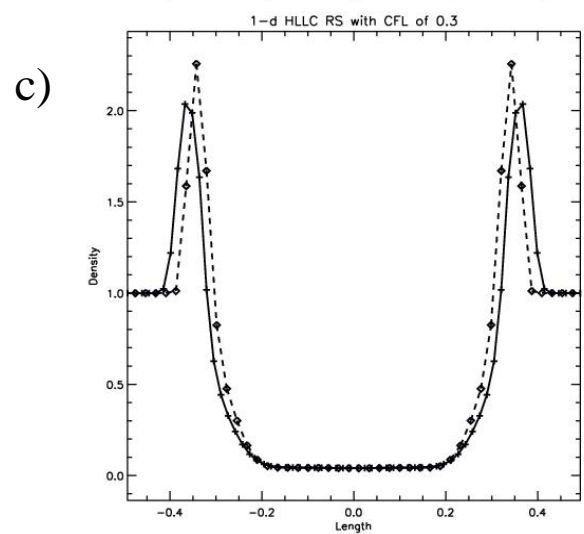 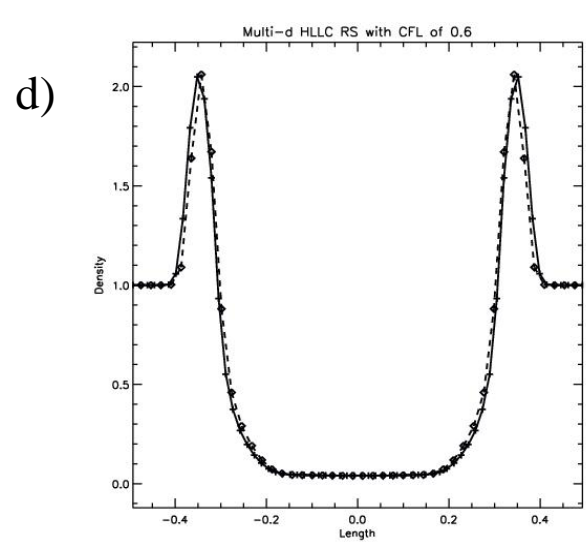

*Figs. 6a shows the density in the midplane of the hydrodynamic blast wave when a conventional second order Godunov scheme was used. Fig. 6b shows the same information when the scheme from the previous section was used. Eight contours with ranges of [ 0.0406, 2.26] and [ 0.0403, 2.06] are shown for Figs. 6a and 6b respectively. An MC limiter was used in both simulations. We also plot out the density along the x-axis (solid line) and along a 45º direction (dashed line). Fig. 6c shows such a plot for the data in Fig. 6a. Fig. 6d shows such a plot for the data in Fig. 6b.*

a) 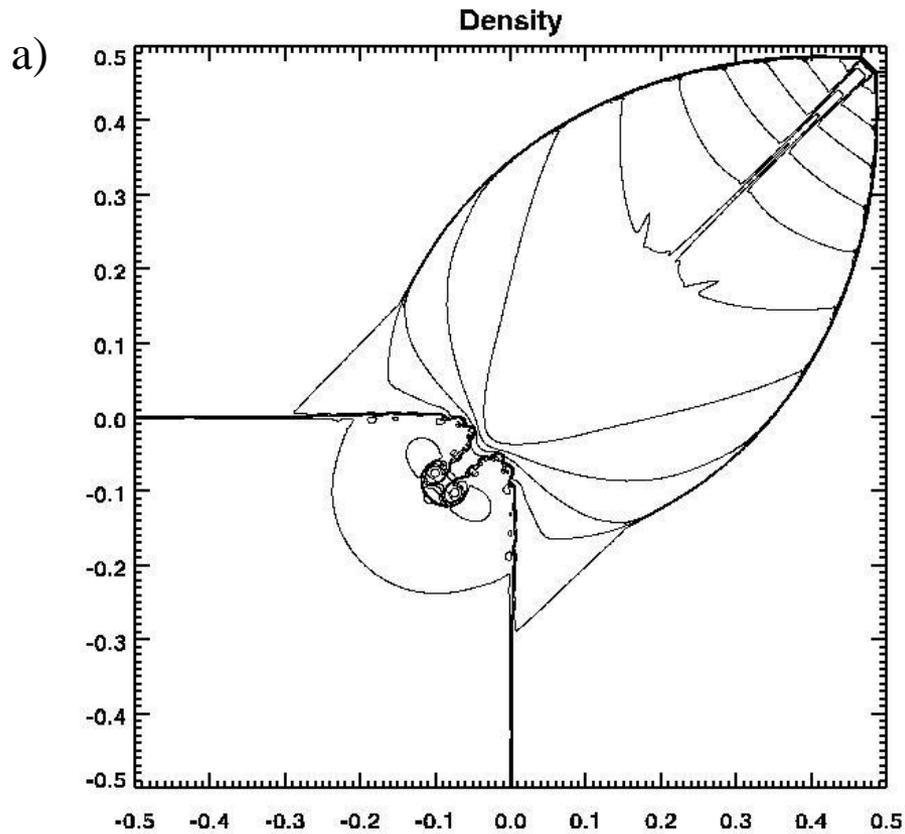 b) 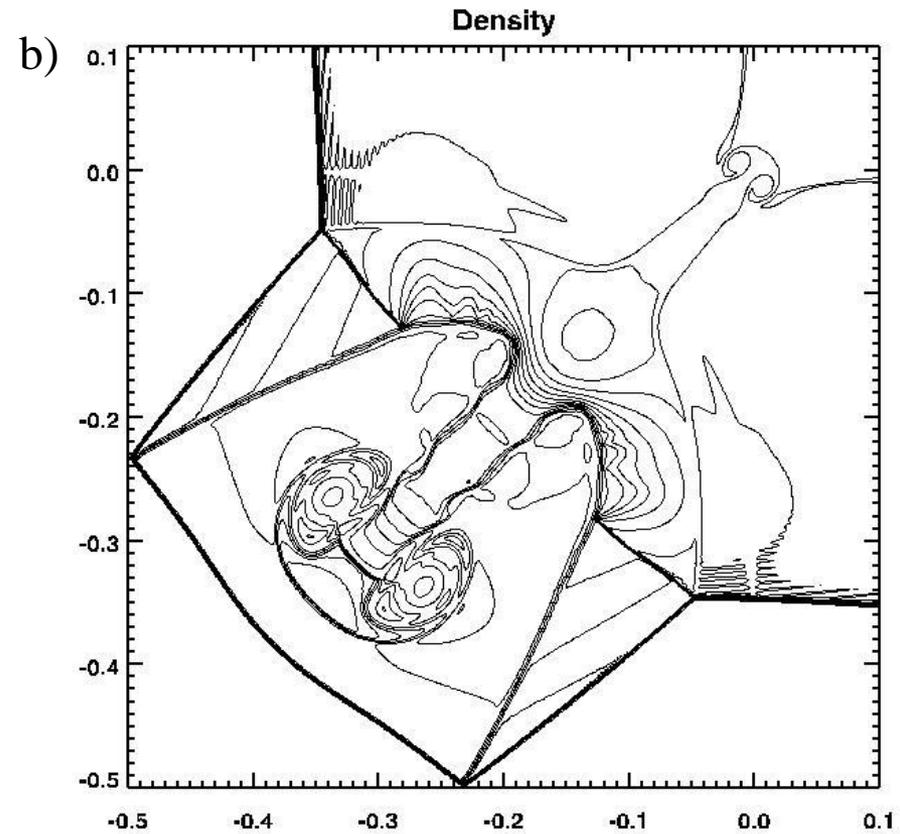

Figs. 7a and 7b show the two density variables from the two-dimensional Riemann problems disccussed in the text. The density variable at the latest time from the first two-dimensional Riemann problem is shown in Fig. 7a. The lower left quadrant of the density variables from the second two-dimensional Riemann problem is shown in Fig. 7b. Twenty contours were used in each figure, with a range of [0.531 , 1.71] for Fig. 7a and [ 0.137, 1.75] for Fig. 7b.

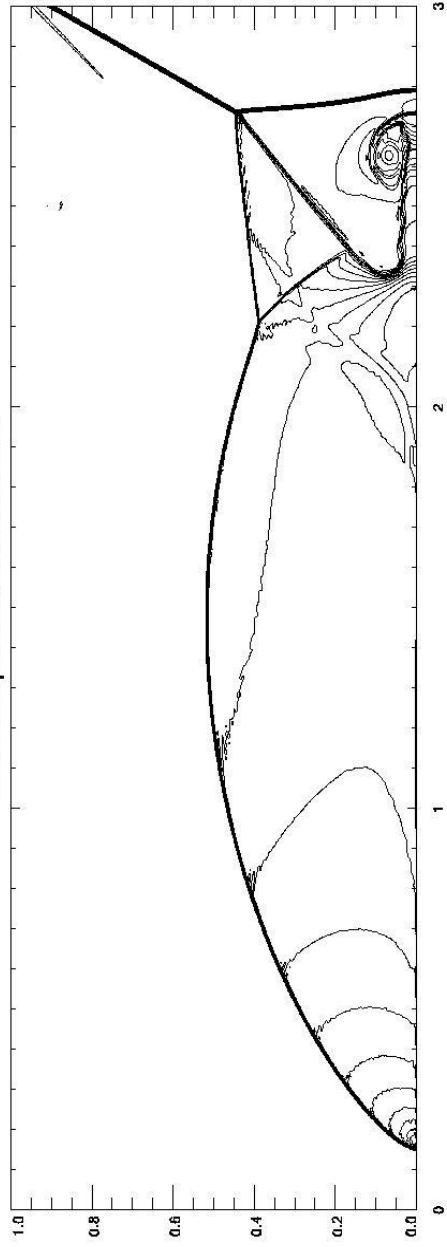

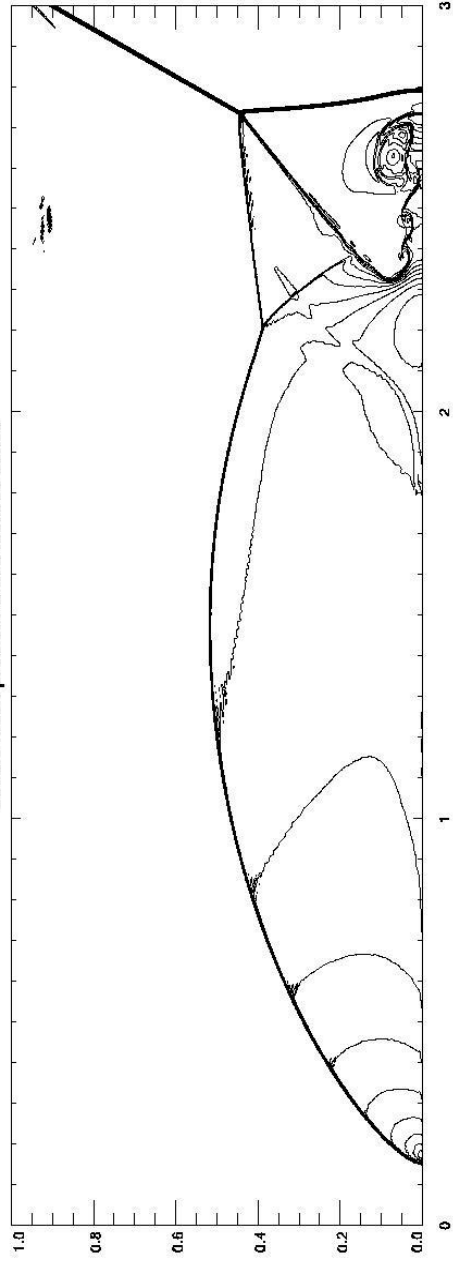

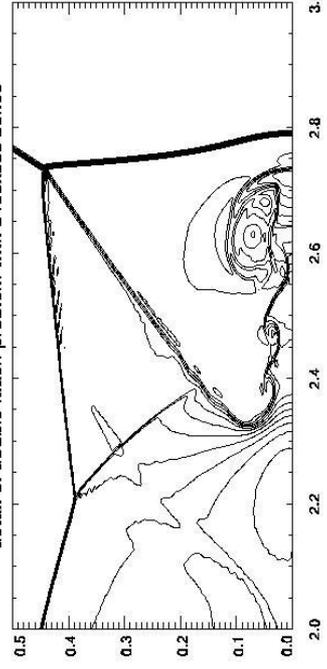

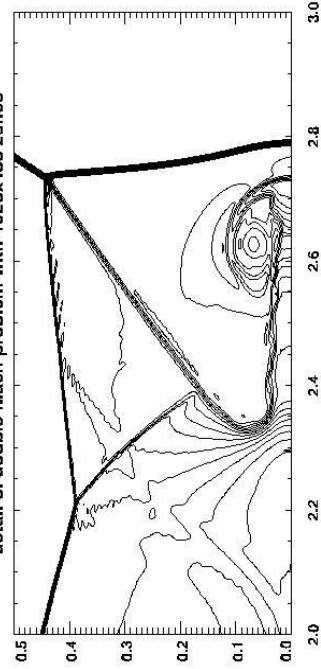

*Figs. 8a and 8b show the density variables at the final time of the double Mach reflection problem at 1920×480 and 2400×600 zone resolution. The two smaller panels at the bottom of the figure, i.e. Figs. 8c and 8d, show a blow-up of the region around the Mach stem from the two former computations. Twenty-five contours with a range of [ 1.77, 22.44] were used.*

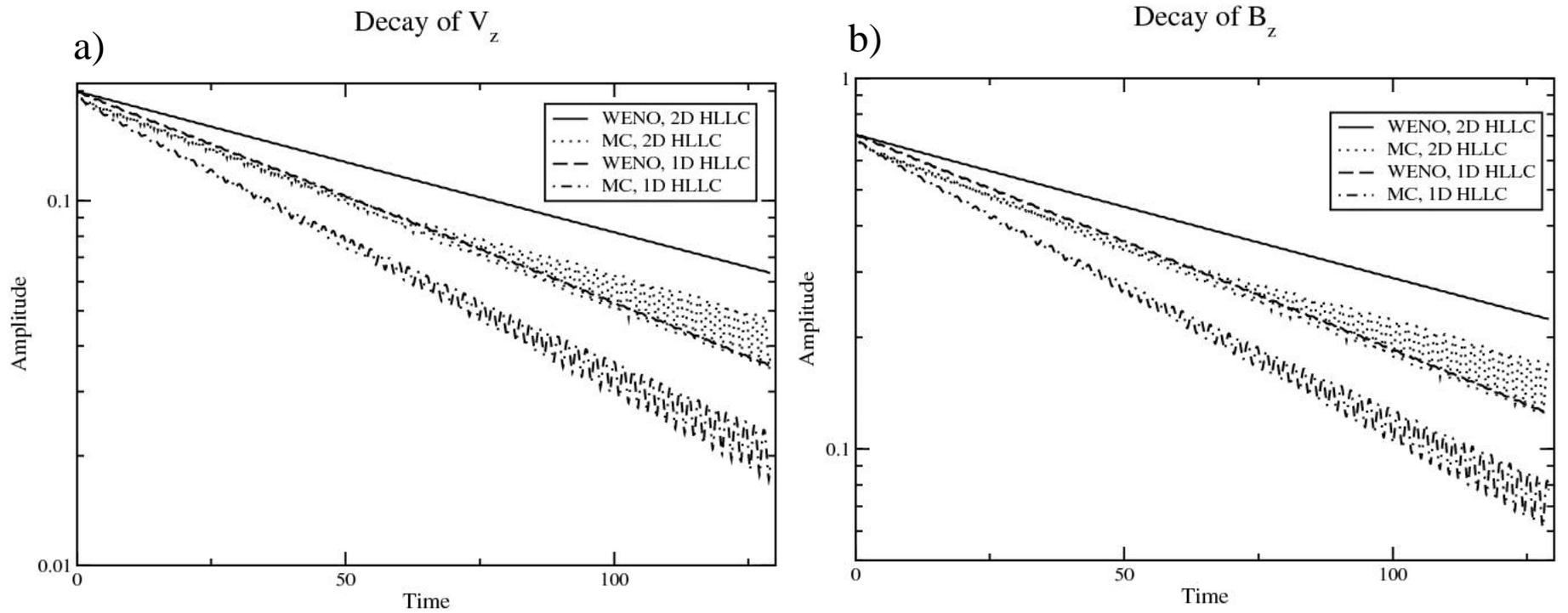

Fig. 9 presents log-linear graphs that show the decay in the amplitude of the Alfven wave with time. Figs. 9a and 9b show the z-velocity and z-magnetic field respectively when different schemes are used. The solid and dotted curves show the decay in the Alfven wave when the multidimensional Riemann solver is used along with the piecewise linear part of an $r=3$ WENO limiter and the MC limiter respectively. The dashed and dot-dash curves show the decay in the Alfven wave when a one-dimensional HLLC Riemann solver is used in a conventional Godunov scheme with the same two reconstruction strategies quoted previously.

a) 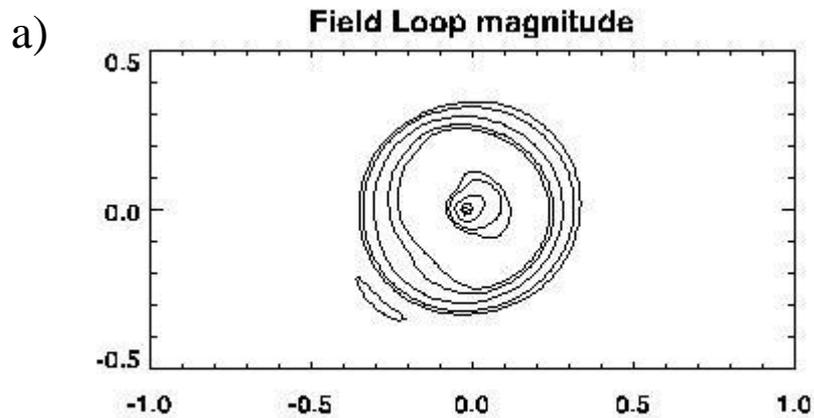 b) 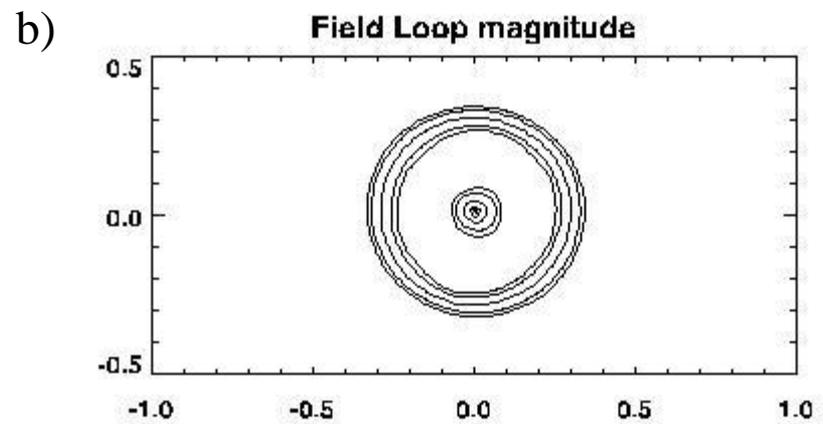

*Figs. 10a and 10b show the magnitudes of the magnetic field for the field loop advection problem after it has executed one complete orbit around the computational domain. Fig. 10a is based on conventional second order Godunov methods with the Gardiner & Stone [27] fix for the electric field. Fig. 10b presents the results from the multidimensional HLLC Riemann solver-based scheme. The field is expected to be uniform in the interior of the loop and zero outside the loop; as a result, four contours were used to demarcate the boundaries of the magnetic flux density.*

a) 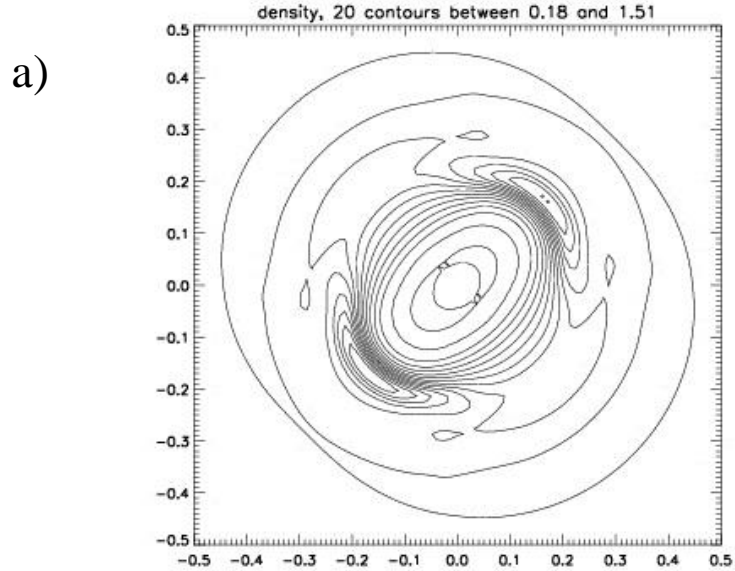
b) 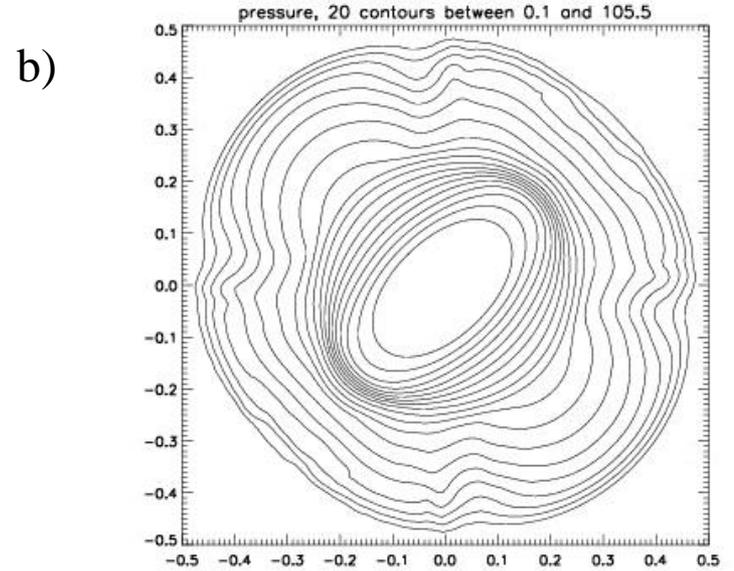
c) 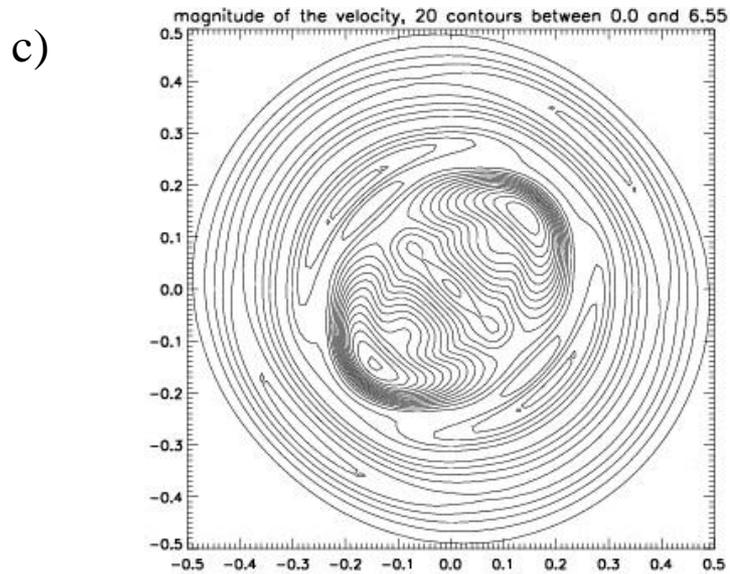
d) 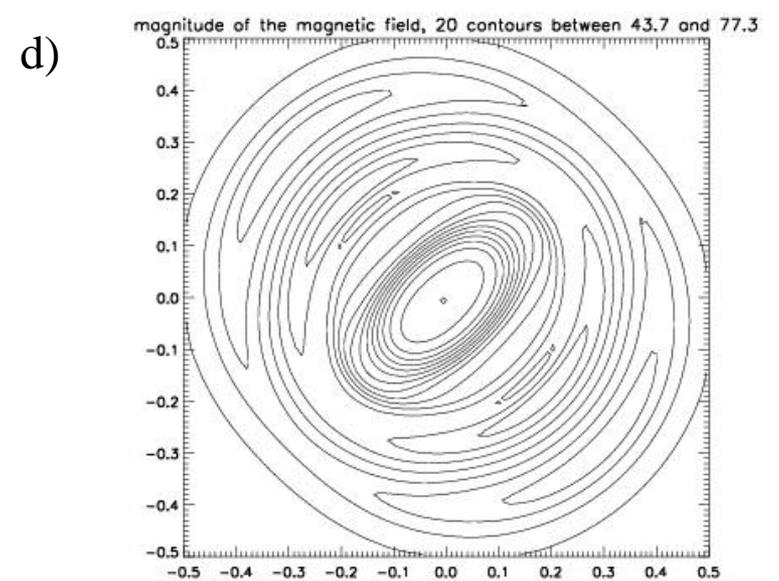

*Figs. 11a through 11d show the density, pressure, magnitude of the velocity and magnitude of the magnetic field respectively in the midplane of the simulation. Each figure shows 20 contours. The density range is [ 0.18, 1.51]; the pressure range is [ 0.1, 105.5]; the range in the velocity is [ 0, 6.55]; the range in the magnetic field is [ 43.7, 77.3].*